\documentclass[superscriptaddress, amsmath,amssymb, aps, pra, twocolumn]{revtex4-1}
\usepackage[
                 colorlinks = true,  
                 linkcolor = blue,
                 urlcolor  = blue,
                 citecolor = blue
                 ]{hyperref} 
\usepackage{enumerate}
\usepackage{tikz}
\usetikzlibrary{patterns}
\usepackage[caption=false,singlelinecheck=off]{subfig}
\usepackage{slashed}
\usepackage{braket}
\usepackage{subfloat}
\usepackage{bbold}
\usepackage{booktabs}
\usepackage{blindtext}
\usepackage{multirow}
\usepackage{bbm}
\usepackage{graphicx}
\usepackage{dcolumn}
\usepackage{bm}
\usepackage[mathlines]{lineno}
\usepackage[normalem]{ulem}

\begin{document}
\title{Conductance plateaus and shot noise in fractional quantum Hall point contacts}
\author{Christian Sp\r{a}nsl\"{a}tt}
\affiliation{Institut f\"{u}r Nanotechnologie, Karlsruhe Institute of Technology, 76021 Karlsruhe, Germany}
\affiliation{Institut f\"{u}r Theorie der Kondensierte Materie, Karlsruhe Institute of Technology, 76128 Karlsruhe, Germany}
\author{Jinhong Park}
\affiliation{Institute for Theoretical Physics, University of Cologne, Z\"{u}lpicher Str. 77, 50937 K\"{o}ln, Germany}
\affiliation{Department of Condensed Matter Physics, Weizmann Institute of Science, Rehovot 76100, Israel}
\author{Yuval Gefen}
\affiliation{Institut f\"{u}r Nanotechnologie, Karlsruhe Institute of Technology, 76021 Karlsruhe, Germany}
\affiliation{Department of Condensed Matter Physics, Weizmann Institute of Science, Rehovot 76100, Israel}
\author{Alexander D. Mirlin}
\affiliation{Institut f\"{u}r Nanotechnologie, Karlsruhe Institute of Technology, 76021 Karlsruhe, Germany}
\affiliation{Institut f\"{u}r Theorie der Kondensierte Materie, Karlsruhe Institute of Technology, 76128 Karlsruhe, Germany}
\affiliation{Petersburg Nuclear Physics Institute, 188300 St. Petersburg, Russia}
\affiliation{L.\,D.~Landau Institute for Theoretical Physics RAS, 119334 Moscow, Russia}

\date{\today}
\begin{abstract}
Quantum point contact devices are indispensable tools for probing the edge structure of the fractional quantum Hall (FQH) states. Recent observations of quantized conductance plateaus accompanied by shot noise in such devices, as well as suppression of Mach-Zehnder interference, call for theoretical explanations. In this paper, we develop a theory of FQH edge state transport through quantum point contacts, which allows for a generic Abelian edge structure and assumes strong equilibration between edge modes (incoherent transport regime).  We find that conductance plateaus are found whenever the quantum point contact locally depletes the Hall bar to a stable region with a filling factor lower than that of the bulk and the resulting edge states equilibrate. The shot noise generated on these plateaus can be classified according to 13 possible combinations of edge charge and heat transport in the device.  We also comment on a relation between the the emergence of quantized plateaus and the suppression of Mach-Zehnder interference. Besides explaining recent experimental findings, our results provide novel insights and perspectives on quantum point contact devices in the FQH regime.
 \end{abstract}
\maketitle
\section{\label{sec:Introduction}Introduction}
The fractional quantum Hall (FQH) effect~\cite{Stormer1982,Laughlin1983} is a paradigmatic realization of the bulk-boundary correspondence: the topological order~\cite{Wen1990a,Wen2004Book} of the two-dimensional (2D), gapped bulk manifests itself holographically on the edge through a gapless one-dimensional topological quantum liquid. This so-called chiral Luttinger liquid, pioneered in several papers by Wen~\cite{Wen1990b,Wen1992,Wen1994,Wen1995} (see also Ref.~\onlinecite{Chang2003} for a review), has its origin in a requirement of the cancellation of the $U(1)$ gauge anomaly in (2+1)D Abelian Chern-Simons theory~\cite{Callan1985,Stone1991,Wen1995}. While the latter topological field theory provides a complete description of the low-energy degrees of freedom in the bulk, it suffers from an anomalous term which violates charge-conservation symmetry in the presence of boundaries. This anomaly is cured by fixing the gauge on the boundary, which naturally introduces the chiral Luttinger liquid hosting the same anomaly as the bulk but with opposite sign. The two anomalous terms therefore cancel exactly and render the full theory consistent and anomaly-free.

The topological properties of a chiral Luttinger liquid describing an Abelian FQH edge are fully specified by the so-called $K$-matrix~\cite{WenZee1992}. This matrix is a topological quantity (i.e. it is fixed up to basis changes for a given topological phase) and carries, in particular, information about the bulk filling factor, $\nu_{\rm{B}} \in \mathbb{Q}$, as well as about charges and exchange statistics of  quasiparticle excitations. In addition, the theory involves non-topological parameters including the edge-channel velocities and inter-channel short-range density-density interactions.

Through the bulk-boundary correspondence,  some experimentally measurable transport characteristics of the edge are quantized under appropriate conditions (discussed below), manifesting the topological order in the bulk. Most prominently, the electrical Hall and two-terminal conductances, $G_H$ and $G$,   
\begin{equation}
G_H = \nu_{\rm{B}} e^2 / h, \qquad G = |G_H|.
\label{eq:G}
\end{equation}
are determined by the filling factor $\nu_{\rm{B}}\equiv\sum_i \delta \nu_i$, where $\delta \nu_i$  are the eigenvalues of $K$ representing the filling-factor discontinuities associated with the edge modes. Furthermore, the thermal Hall and two-terminal conductances, $G^Q_H$ and $G^Q$, 
\begin{equation}
G^Q_H = \nu_Q\kappa T, \qquad G^Q = |G^Q_H|,
\label{eq:GQ}
\end{equation}
manifest $\nu_Q\equiv \sum_i{\rm sgn}(\delta \nu_i) \equiv n_d-n_u\in \mathbb{Z}$, the difference between the numbers of ``downstream'' , $n_d$, and ``upstream'', $n_u$,  
edge channels~\cite{Kane1997,Capelli2002}. Here, the ``downstream''  direction is that of predominant charge propagation as set by the magnetic field, and ``upstream'' is the direction opposite to ``downstream''. 
Further, $T$ is the temperature, $\kappa = \pi^2 k_B^2/3h$, and $k_B$ is Boltzmann's constant. 

The remarkable quantizations in Eqs.~\eqref{eq:G} and \eqref{eq:GQ} hold very generally for quantum Hall states with ``maximally chiral'' edges such as, e.g., the Laughlin states~\cite{Laughlin1983} or the integer quantum Hall states~\cite{Klitzing1980,Buttiker1988}. Such edges support channels which all propagate unidirectionally, i.e., all eigenvalues of the $K$ matrix have the same sign (strictly speaking, this holds under assumption that no edge reconstruction~\cite{Meir1994,Wang2013} takes place). In this situation, the transport coefficients $\nu_{\rm{B}}$ and $\nu_Q$ are extremely robust against disorder due to the absence of backscattering.

Edges of generic FQHE states are not maximally chiral, i.e., they consist of counterpropagating (downstream and upstream) modes. Equivalently, the $K$ matrix of such a state has eigenvalues of both signs. In particular, edges of hole-conjugate states (i.e. filling factors $1/2<\nu_{\rm B}<1$) are of this type, as was theoretically predicted~\cite{Wen1992, Wen1995} and experimentally verified~\cite{Bid2010, Venkatachalam2012, Altimiras2012, Inoue2014}. A prominent example is the $\nu_{\rm B} = 2/3$ state which possesses 
one downstream edge channel associated with filling factor discontinuity $\delta \nu = 1$ and one upstream channel with $\delta \nu = -1/3$ \cite{MacDonald1990,Wen1992}. 
The renormalization-group analysis of a disordered interacting 2/3 edge was developed by Kane, Fisher, and Polchinski (KFP) \cite{Kane1994}. 

In general, the topological quantizations \eqref{eq:G} and \eqref{eq:GQ} do not hold for edges with counterpropagating modes since the conductances depend on degree of backscattering. However, the quantizations \eqref{eq:G} and \eqref{eq:GQ}  are restored in the incoherent (fully equilibrated) regime~\cite{Kane1995, Kane1997, Protopopov2017,Nosiglia2018}.  The incoherent regime takes place in the limit $l_{\rm eq}\ll L$, where $L$ is the edge length and  $l_{\rm eq}$ is a characteristic length for inelastic equilibration. Beyond $l_{\rm eq}$, individual channels fully equilibrate and form hydrodynamic modes. What remains of the underlying edge structure is the transport coefficients $\nu_{\rm{B}}$ and $\nu_Q$ which are entirely dictated by the topological order of the bulk. 

The incoherent regime has been discussed in early theoretical papers~\cite{Kane1995,Kane1997} and later in the context of line junctions~\cite{Sen2008,Rosenow2010}.
More recently, Refs.~\onlinecite{Protopopov2017,Nosiglia2018,Aharon2019,Park2019} performed a systematic study of the electric and thermal transport for 
the incoherent $\nu_{\rm B}=2/3$ edge. Further, Ref.~\onlinecite{Protopopov2017} reported the analysis of the crossover between the coherent ($L \ll l_{\rm eq}$) and incoherent ($L \gg l_{\rm eq}$) regimes with increasing temperature (or, equivalently, increasing sample length). This analysis demonstrates that the quantizations \eqref{eq:G} and \eqref{eq:GQ} are unique features of the incoherent regime. To date, this quantization holds in nearly all FQH experiments on complex edges, which implies that the experimentally studied systems are generically in the incoherent regime. Reaching the coherent regime in conventional FQH structures would require either extremely low temperatures or very small distances between contacts. Only very recently was the full coherent-to-incoherent crossover (as predicted in Ref.~\onlinecite{Protopopov2017}) experimentally observed in a specially designed double-well structure that permitted a high degree of control over the intermode tunneling~\cite{Cohen2019}. 

Measurement of thermal transport characteristics is much more difficult than electric measurements and requires substantially more sophisticated schemes. While the thermal conductance measurements~\cite{Jezouin2013,Banerjee2017,Banerjee2018, Srivastaveaaw5798} are  in agreement with the incoherent result \eqref{eq:GQ}, only the two-terminal conductance $G^Q$ was determined, which is proportional to the absolute value $|\nu_Q|$ of the topological invariant. Such measurements do not provide the information about the sign of $\nu_Q$ which distinguishes between downstream and upstream heat propagation. Motivated by this limitation, a complementary and fully electrical approach to identify upstream heat propagation with shot noise measurements was recently proposed in Refs.~\onlinecite{Park2019,Spanslatt2019}. There, it was shown that in the incoherent regime, the noise falls into three topologically distinct universality classes depending on the direction of heat propagation with respect to that of the charge flow (defined as downstream).  In particular, when $\nu_{Q} < 0$, the noise is constant as a function of the edge length up to exponentially small corrections in $L/l_{\rm eq}$.

 Another powerful tool for probing the FQH edge structure is the quantum point contact (QPC) geometry~\cite{Milliken1996,Roddaro2004,Roddaro2004b,Miller2007,Radu2008, Hong2017}. By using appropriately etched gates that locally deplete the 2D electron gas, local constrictions in the FQH sample are formed. When the gate voltage varies, the size of the depletion region, and thereby the sample boundary, is modified. The two edges of the sample can therefore be brought into proximity which allows inter-edge tunneling and thus leads to a gradual change of the conductance through the sample. 
 
 QPC experiments performed for the hole-conjugate states~\cite{Ando1998,Bid2009, Bhattacharyya2019} have brought forward a number of remarkable observations. First, it was found that, as the gate voltage is tuned such that the QPC varies from fully open to fully closed, the resulting two-terminal conductance, $G$, typically develops a set of quantized plateaus. The most prominent is the plateau with $G = (1/3) e^2 / h$; other  plateaus that have been reported are characterized by $G$ equal to 1/5, 4/5, 2/3, 2/5, and 3/7 in units of $e^2/h$.  Secondly, these plateaus are characterized by shot noise $S$ that was measured in Ref.~\onlinecite{Bhattacharyya2019} for the 1/3 plateau for bulk filling factors $\nu_{\rm{B}}= 1$, $\nu_{\rm{B}}= 2/3$, $\nu_{\rm{B}}=3/5$, and $\nu_{\rm{B}}=4/7$. The strength of this noise was characterized by the Fano factor that was defined according to 
 \begin{equation}
\label{eq:FanoFactor}
	S = 2FeI_{\rm imp}\tau(1-\tau),
\end{equation}
where $I_{\rm imp}$ is the current impinging on the QPC and $\tau$ is the transmission through the QPC (defined as the ratio of the current through to QPC and $I_{\rm imp}$). 
The found values of $F$ were of order unity, implying that the noise is strong. Interestingly, the obtained values of $F$ were always close to the bulk filling factor $\nu_{\rm{B}}$. A qualitatively similar behavior, with noisy intermediate conductance plateaus, was found in a double-QPC geometry~\cite{Sabo2017}.

Complementary experimental information on the coherence in the system was obtained from studying an electronic Mach-Zehnder interferometer defined by two QPCs~\cite{Bhattacharyya2019}. The visibilities of the interference patterns were found to be quite clear for $\nu_{\rm{B}} \geq 5/3$ but were drastically washed out for lower filling factors. This ``melting of interference'' was observed to be correlated with the onset of the $G=(1/3)e^2/h$ plateau in transport through the QPC. The authors of Ref.~\onlinecite{Bhattacharyya2019}. attributed the loss of interference to proliferation of neutral modes~\cite{Park2015,Goldstein2016} induced by edge reconstruction due to a soft edge confinement potential~\cite{Meir1994,Wang2013, Sabo2017}. 

The onset of the 1/3 plateau for the $\nu_B=2/3$ state is consistent with a picture developed by Wang, Meir, and Gefen (WMG)~\cite{Meir1994,Wang2013}. The starting point of the WMG theory is the edge reconstruction of the 2/3 edge due to a sufficiently small slope of the edge confinement potential. The corresponding edge structure includes, in addition to $\delta \nu = 1$ and $\delta \nu = - 1/3$ modes of the ``conventional'' 2/3 edge,
two counterpropagating modes with $\delta \nu = \pm 1/3$.   WMG demonstrated that, 
under certain assumptions, disorder can drive such an edge towards a renormalization-group fixed point  with two downstream $\delta \nu = 1/3$ charge modes and two upstream neutral modes. At this fixed point, the presence of the $G=(1/3)e^2/h$ plateau is accounted for by assuming that 
in certain ranges of the gate voltage controlling the QPC, the outermost charged channel is transmitted through the QPC while 
the inner charged one is reflected. The noise on  this plateau is further attributed to the generation of quasiparticle-quasihole pairs created by the equilibration-induced decay of the neutral modes~\cite{Sabo2017}. 

Even though this theory seems to successfully explain the experiments for $\nu_{\rm B}=2/3$, there are two important points of critique. 
First of all, sharp plateaus of $G = (1/3) e^2 / h $ were found also for $\nu_{\rm{B}}=3/5$, $\nu_{\rm{B}}=4/7$, and perhaps most surprisingly, for the integer state $\nu_{\rm{B}}=1$~\cite{Bhattacharyya2019}. Additional plateaus, albeit less prominent, at $G=(2/5)e^2/h$ (for both $\nu_{\rm{B}}=3/5$ and $\nu_{\rm{B}}=4/7$)
and $G=(3/7)e^2/h$ (only for $\nu_{\rm{B}}=4/7$) were observed as well. It seems difficult to imagine that edges of all hole-conjugate states, as well as $\nu_{\rm{B}}=1$ edge, would get reconstructed by additional $\delta\nu = \pm 1/3$ modes, and we are not aware of any microscopic analysis that would support this feature. It is also difficult to imagine that an edge would undergo such a complex edge reconstruction that it would explain multiple plateaus. Secondly,  the WMG theory is based on assumption that the renormalization-group flow has enough room to drive the system to the fixed point with coherent neutral modes (in analogy with the KFP fixed point \cite{Kane1994}). At the same time, 
current FQH experiments appear to be far from the transport regime involving coherent neutral modes (see Ref.~\onlinecite{Protopopov2017} for a detailed analysis). Specifically, reaching such a regime  would require much lower temperatures than in current experiments (inelastic processes at finite temperature stop the renormalization-group flow). The hallmark of such a regime---strong mesoscopic fluctuations of $G$---has never been observed experimentally. 
\begin{figure}[t]
\captionsetup[subfigure]{position=top,justification=raggedright}
\subfloat[]{
\includegraphics[width=0.95\columnwidth]{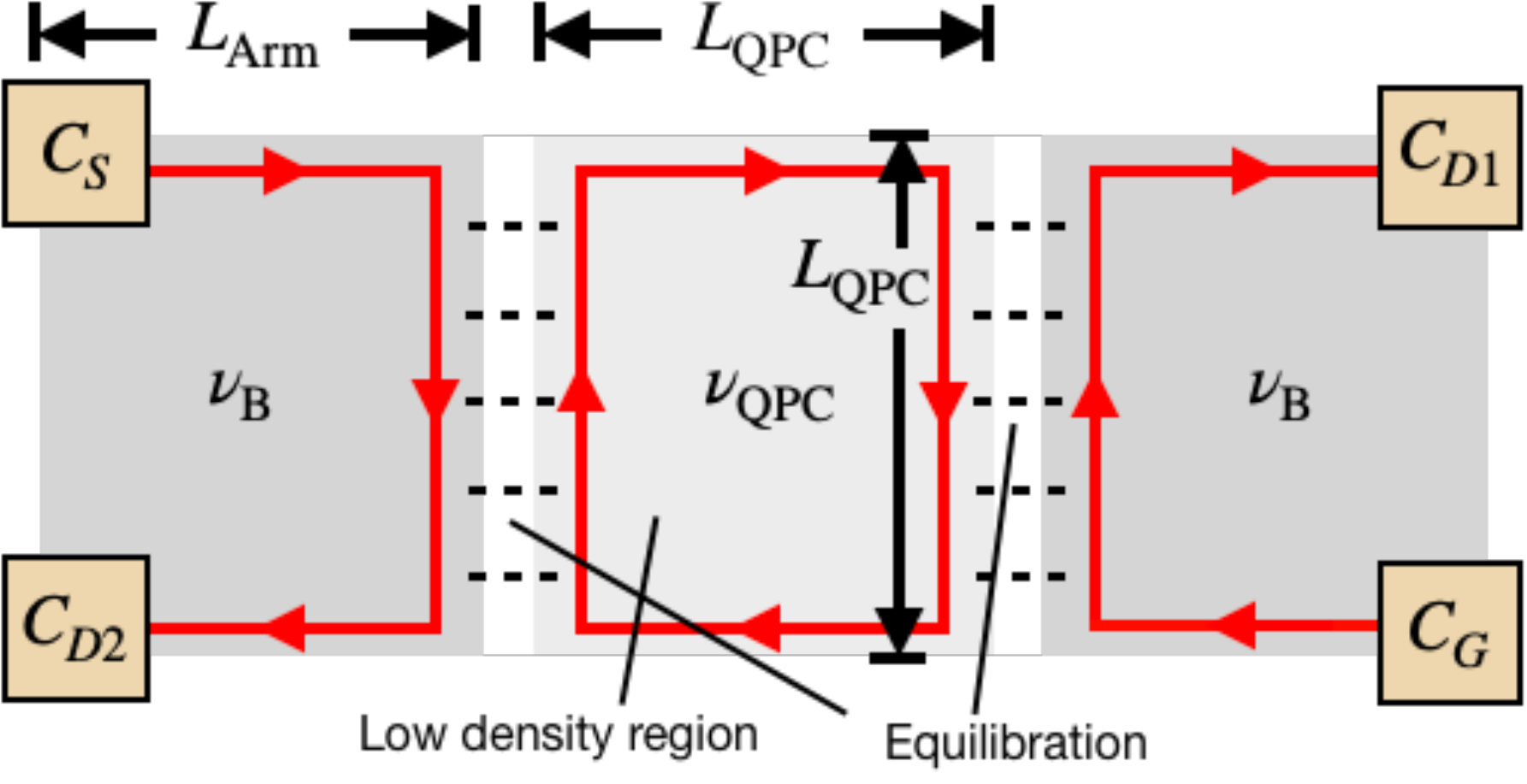}
\label{fig:QPCSetup}}
\\[0.0cm]
\subfloat[]{
\includegraphics[width =0.95\columnwidth]{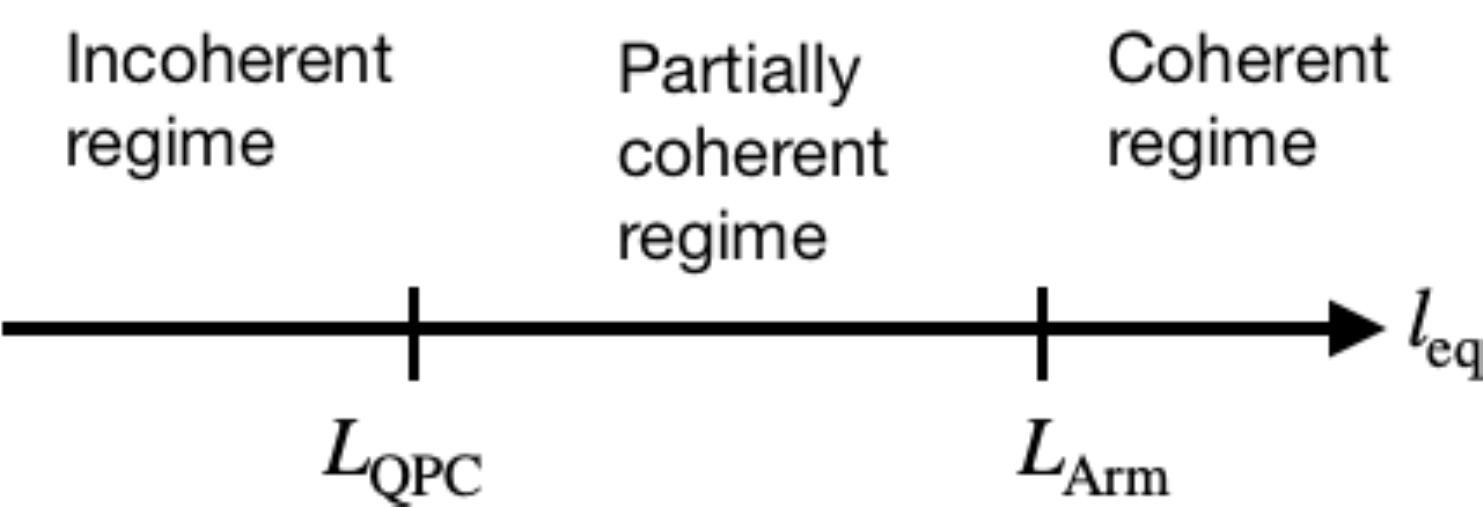}
\label{fig:LengthScales}}
\caption{(a) Model of a FQH QPC device: in a FQH state with filling factor $\nu_{\rm B}$, a low density region with filling factor $\nu_{\rm QPC}<\nu_{\rm B}$ is formed by the QPC gate. The edge states of these regions (with net chirality depicted by the red arrows) interact and equilibrate in two narrow regions (ladders of black dashed lines).  An edge current injected from the source contact $C_S$ reaches, after equilibration with the low density edge, the two drains $C_{D1}$ and $C_{D2}$. The contact $C_G$ is without loss of generality taken to be grounded. (b) Hierarchy of length scales in the QPC device. 
The total device length $L_{\rm Arm}$ is assumed to be much larger than the length scale of the QPC: $L_{\rm QPC}$ (this condition is always fulfilled experimentally).
Depending on the relations between the inelastic equilibration length $l_{\rm eq}$ and $L_{\rm Arm}$ and $L_{\rm QPC}$, we distinguish between three transport regimes. In this paper we focus primarily on the incoherent regime $l_{\rm eq}\ll L_{\rm QPC}\ll L_{\rm Arm}$, but we will also briefly comment on the partially coherent regime $L_{\rm QPC}\ll l_{\rm eq} \ll L_{\rm Arm}$ towards the end of the paper. The fully coherent regime $L_{\rm QPC}\ll L_{\rm Arm}\ll l_{\rm eq}$ (which may only be achieved at extremely low temperatures or in devices with a special control over $l_{\rm eq}$) is not considered in this work.}
\label{fig:IntroFig}
\end{figure}
The goal of this paper is to develop a theory of transport of FQH edge states through QPCs applicable to generic bulk filling factors $\nu_{\rm B}$, which would explain the emergence of multiple noisy plateaus and predict the associated transport characteristics. For this purpose, we model the QPC as a low density constriction with filling factor $\nu_{\rm QPC}<\nu_{\rm B}$ (see Fig.~\ref{fig:QPCSetup}). Similar models were considered earlier in Refs.~\onlinecite{Beenakker1990,Rosenow2010}. Our main focus here is the incoherent regime, where the equilibration length $l_{\rm{eq}}$ is much smaller than both $L_{\rm{QPC}}$ and $L_{\rm{Arm}}$ (see Fig.~\ref{fig:LengthScales}). As indicated in the above discussion, the motivation for focusing on the incoherent regime is twofold: (i) the FQH edge transport is incoherent in almost all experiments; (ii) the incoherent regime has a much higher degree of universality. We study the electrical and thermal transport characteristics, including the conductances and the noise.
 Our most salient findings are as follows:
\begin{enumerate}[i)]
	\item We show that the considered model accounts for all observations of conductance plateaus (Sec.~\ref{sec:SQPC results}). The injected current impinging onto the QPC, $I_{\rm imp}=\nu_{\rm B}e^2V_0/h$ (where $V_0$ is the bias voltage),  splits into two parts which eventually reach two different drains $C_{D1}$ and $C_{D2}$ (see Fig.~\ref{fig:QPCSetup}).  The conductances measured in these drains in the incoherent regime (full equlibration) are $G_{D1}=(\nu_{\rm B}-\nu_{\rm QPC})e^2/h$ and $G_{D2}=\nu_{\rm QPC}e^2/h$. Remarkably, $G_{D2}$ depends only on $\nu_{\rm QPC}$. Assuming that the FQH state corresponding to the filling $\nu_{\rm QPC}$ is stable in a certain range of the densities, the observation of conductance plateaus follows. 
Since $\nu_{\rm QPC}=1/3$ is the most stable state in the whole hierarchy of FQH states, the $G=(1/3)e^2/h$ plateau is the most visible one.

	\item We show that the experimental results for double-QPC geometries~\cite{Sabo2017} also can be accounted for by our model (Sec.~\ref{sec:DQPCresults}). With a short distance between the two QPCs,  the setup is found to be equivalent to a single QPC (see Fig.~\ref{fig:DQPCExpa}), leading to $G_{D1}=\nu_{\rm QPC}e^2/h$. With more distant QPCs, the setup is instead equivalent to two QPCs in series (see Figs.~\ref{fig:DQPCb} and~\ref{fig:DQPCExpb}) and we find $G_{D1}=(\nu_{\rm QPC}^2/ \nu_{\rm B})e^2/h$. For the specific configuration $\nu_{\rm B} = 2/3$ and $\nu_{\rm QPC} = 1/3$, the conductance changes from $G_{D1} = e^2/ 3h$ to $G_{D1} = e^2/ 6h$ as the distance between the QPCs increases, which is in excellent agreement with experiment. 

	\item We derive the topological characteristics of the shot noise for the possible combinations of electrical and thermal transport in the single-QPC geometry (Sec.~\ref{sec:QuantNoise}). We find that the noise falls into $13$ topologically distinct classes (see Tab.~\ref{tab:QPCQualitative}). The qualitative asymptotic length dependencies of the noise on $L_{\rm{QPC}}$ and $L_{\rm Arm}$ are only governed by the directions of heat propagation along the edge segments forming the device. Most interestingly, some of the classes exhibit super-Poissonian noise: $F>1$.

	\item We compute quantitatively the noise on the $G=(1/3)e^2/h$ plateau for a few representative FQH states (see Sec.~\ref{sec:QualNoise}). No quantization of the corresponding Fano factors is found.

	\item Our theory also explains the strong suppression of the Mach-Zehnder interference in devices that show the $1/3$ plateau for the QPC transport.
	 Indeed, the emergence of the (most prominent) $1/3$ conductance  plateau in our theory is a result of the incoherent character of the transport, $l_{\rm{eq}} \ll L_{\rm{QPC}}$. The inelastic processes establishing this incoherent transport regime will also destroy the coherence of Mach-Zehnder interferometry. The suppression of the Mach-Zehnder interference pattern is thus not related to neutral modes but rather has a much more general origin:  it is an indicator of the incoherent regime. 
\end{enumerate}
The remainder of this paper is organized as follows. In Sec.~\ref{sec:Model}, we present our model of the incoherent regime and derive transport equations determining the local voltages, local temperatures, and shot  noise along a FQH edge. We then apply the model to QPC geometries, obtaining expressions of the conductance in Sec.~\ref{sec:QPCApplications} and the noise in Sec.~\ref{sec:NoiseQPCs}. We discuss our results in Sec.~\ref{sec:Discussion} and conclude our studies with a summary and an outlook in Sec.~\ref{sec:Summary}.

\section{\label{sec:Model}Model of the incoherent regime}
\begin{figure}[t]
\includegraphics[width=0.95\columnwidth] {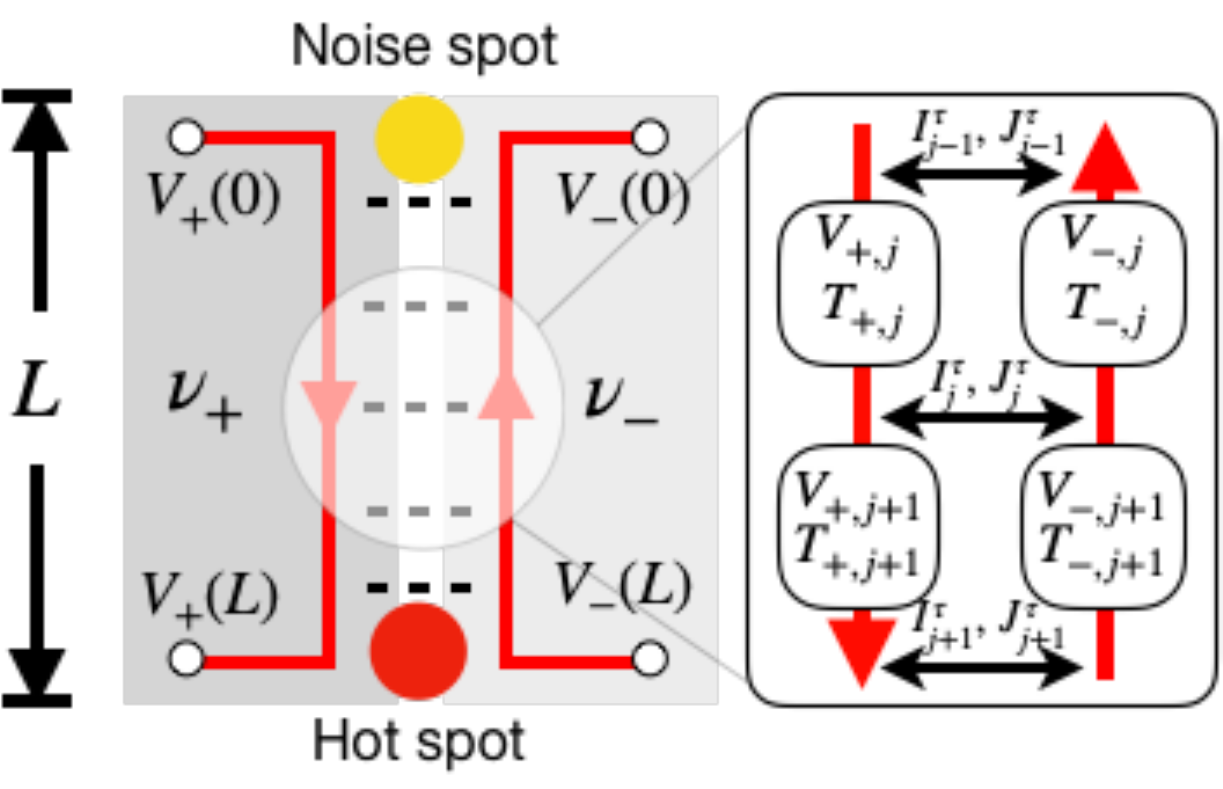}\caption{\label{fig:EdgeSegment} Effective model of a FQH edge segment (either a complex edge or a line junction of two edges) in the incoherent regime. Two hydrodynamic modes with filling factors $\nu_+$ and $\nu_-$ equilibrate due to random tunneling. The local voltages $V_{\pm,j}$ and temperatures $T_{\pm,j}$ of the modes are determined by local virtual reservoirs (labelled by $j$) which drive charge $I^{\tau}_{j}$ and energy $J^{\tau}_{j}$  tunneling currents (the reservoirs do not absorb charge or energy). Given that $\nu_+>\nu_-$, the equilibration results in a voltage drop and heat generation at the \textit{hot spot} $\sim x=L$ (red dot). The type of heat conduction is fixed by the topological quantity $\nu_Q \equiv n_d-n_u$, i.e., the difference between the number of downstream and upstream microscopic channels. If heat reaches the region close to $x=0$, the \textit{noise spot} (yellow dot), shot noise is generated. The locations of the hot and noise spots as well as the value of the noise are independent of the direction in which the voltage bias is put.}
\end{figure}

We consider a general Abelian FQH edge segment (see Fig.~\ref{fig:EdgeSegment}) of length $L$ hosting $n_d$ downstream and $n_u$ upstream propagating channels. The total number of channels is denoted $N\equiv n_d+n_u$. This model describes both a complex edge of a FQH state as well as an interface between two FQH states, i.e., a line junction of the corresponding FQH edges. 

To simplify our treatment of the edge structure, we shall hereafter assume that all edge channels with a given chirality equilibrate into effective hydrodynamic modes on the length scale $\sim a$ which thus serves as the UV cutoff of the model (see Ref.~\onlinecite{Spanslatt2019} for a detailed discussion). The resulting two modes, labelled by $n=\pm$, have effective filling factor discontinuities
\begin{subequations}
	\begin{align}
		\nu_+ = &\;\; \sum_{n=1}^{n_d} \;\;\;\delta \nu_n, \\
		\nu_- = &\sum_{n=n_d+1}^{N} \delta \nu_n,
	\end{align}
\end{subequations}
 respectively. Here, $\delta\nu_n$ are the  filling factor discontinuities of individual channels. Without loss of generality, we hereafter assume $\nu_+>\nu_-$. 
 In the sequel, we will denote the hydrodynamic mode chiralities by $\chi_\pm=\pm 1$. 

To model local equilibration between the two remaining (counterpropagating) hydrodynamic modes, we introduce $M$ virtual reservoirs (which absorb neither charge nor energy) for each mode~\cite{Engquist1981,Nosiglia2018,Park2019,Spanslatt2019}. Since we are only interested in steady state properties and zero frequency noise, we neglect any temporary charge or heat accumulation in the reservoirs. The charge and energy currents along the edge segment are then locally conserved and we can write
\begin{subequations}
\label{eq:ConservationReservoirs}
\begin{align}
& I_{n,j,\textrm{out}} = I_{n,j, \textrm{in}} \equiv I_{n,j}, \\
& J_{n,j,\textrm{out}} = J_{n,j, \textrm{in}} \equiv J_{n,j},
\end{align}
\end{subequations}
where $I(J)_{n,j, \textrm{out}( \textrm{in})}$ is the outgoing (incoming) charge (energy) current of mode $n$ into its reservoir at location $j$. 
Here and below, the quantities are understood as time-averaged (we do not indicate time-averaging explicitly in order to simplify the notation).  
 The local voltages 
 \begin{equation}
 \label{eq:voltage-current}
 V_{n,j}\equiv \frac{h}{e^2  \nu_n} I_{n,j}
 \end{equation}
 and temperatures $T_{n,j}$ drive local charge and energy tunneling currents $I^\tau_{j}$ and $J^\tau_{j}$ at position $j$,
\begin{subequations}
\label{eq:ConservationTunneling}
\begin{align}
&I_{+,j+1} = I_{+,j} - I^\tau_{j},\\
&I_{-,j+1} = I_{-,j} + I^\tau_{j},\\
&J_{+,j+1} = J_{+,j} - J^\tau_{j},\\
&J_{-,j+1} = J_{-,j} + J^\tau_{j}.
\end{align}
\end{subequations}
To lowest order in the dimensionless tunneling coupling $g\ll1$, these currents can be written as
\begin{subequations}
\label{eq:TunnelingCurrents}
\begin{align}
& I^\tau_{j}=g\frac{e^2}{h}\left(V_{+,j}-V_{-,j}\right),\\
& J^\tau_{j}=g\frac{e^2}{2h}\left(V_{+,j}^2-V^2_{-,j}\right) +\gamma g\frac{\kappa}{2} \left(T^2_{+,j}-T^2_{-,j}\right),
\end{align}
\end{subequations}
where $\gamma$ is a parameter of order unity, characterizing the deviation of the ratio of intermode charge and heat tunneling conductances from Wiedemann-Franz law (Wiedemann-Franz law corresponds to $\gamma=1$).
Generally, $\gamma$ depends on the edge structure and the intermode interactions. In particular, for two non-interacting channels with $\delta\nu_1 = 1$ and $\delta\nu_2=\nu = 1/(2p+1)$ with an integer $p$, the value of $\gamma$ was found to be $\gamma=3/(2\nu+1)$~\cite{Nosiglia2018}. For simplicity, we neglect any voltage and temperature dependence in $g$.

By combining Eqs.~\eqref{eq:ConservationTunneling} and~\eqref{eq:TunnelingCurrents} we next derive continuum equations for the voltage, current, temperature, and noise profiles along the edge segment. 

\subsection{\label{sec:VoltageEquilibration} Voltage equilibration and two-terminal charge conductance}
The distance between successive reservoirs is $a$ (the UV cutoff of our model), and we define $x=ja$. The continuum limit is obtained by setting $g\rightarrow 0$, $a\rightarrow 0$  but keeping $x$ and $l \equiv a/g$ constant. We then obtain the following differential equation for the local voltages
\begin{equation}
\label{eq:VoltageEquation}
\partial_{x}\vec{V}(x) = \mathcal{M}_V \vec{V}(x),
\end{equation}
where $\vec{V}(x)=(V_+(x),V_-(x))^T$ (superscript $T$ denotes transposition) and 
\begin{equation}
\label{eq:V2times2}
	\mathcal{M}_V = \frac{1}{l} \begin{pmatrix}
-\chi_+/\nu_+ & \chi_+/\nu_+\\
\chi_-/\nu_- & -\chi_-/\nu_-
\end{pmatrix}.
\end{equation}
The corresponding local electric currents $\vec{I}(x)=(I_+(x),I_-(x))^T$ obey a similar equation
\begin{equation}
\label{eq:CurrentEquation}
\partial_{x}\vec{I}(x) = \mathcal{M}_I \vec{I}(x), \qquad \mathcal{M}_I =   \mathcal{D} \mathcal{M}_V  \mathcal{D}^{-1},
\end{equation}
 with  $\mathcal{D} = \text{diag}(\chi_+ \nu_+,\chi_- \nu_- )$.

We are interested in the solution of Eq.~\eqref{eq:VoltageEquation}  for the case of counterpropagating modes, $\chi_+ = 1$ and $\chi_- = -1$. The equation should be then supplemented by the boundary conditions fixing $V_+(0)$ and $V_-(L)$. We first choose the boundary condition describing the bias applied at the $x=0$ end of the edge segment,  $V_+(0)=V_0$ and $V_-(L)=0$, which corresponds to the downstream direction of the electric current. The solution of Eq.~\eqref{eq:VoltageEquation} then becomes
\begin{subequations}
\label{eq:VProfiles}
	\begin{align}
		& V_+(x) = V_0\frac{\nu_+ e^{L/l_{\textrm{eq}}}-\nu_- e^{x/l_{\textrm{eq}}}}{\nu_+e^{L/l_{\textrm{eq}}} - \nu_-}, \\
		& V_-(x) = V_0 \frac{\nu_+ e^{x/l_{\textrm{eq}}}-\nu_+ e^{L/l_{\textrm{eq}}}}{\nu_+e^{L/l_{\textrm{eq}}} - \nu_-},
	\end{align}
\end{subequations}
where we have defined the phenomenological equilibration length $l_{\textrm{eq}} \equiv l\nu_+\nu_-/(\nu_+-\nu_-)$. 
 
 In the incoherent limit, $l_{\textrm{eq}}\ll L$, we obtain $V_{+}(L)\simeq V_0(\nu_+-\nu_-)/\nu_+$ and the ``downstream conductance''  
 \begin{equation}
 \label{eq:Gd}
 G_d \equiv I_+(L)/V_{0}\simeq(\nu_+-\nu_-) e^2/h\equiv\nu e^2/h. 
 \end{equation}
 With reversed boundary conditions, $V_+(0)=0$ and $V_-(L)=V_0$, which corresponds to upstream direction of the electric current, we obtain analogously
\begin{subequations}
\label{eq:VProfiles2}
	\begin{align}
		& V_+(x) = V_0\frac{\nu_-e^{x/l_{\textrm{eq}}}-\nu_-}{\nu_+e^{L/l_{\textrm{eq}}} - \nu_-}, \\
		& V_-(x) = V_0 \frac{\nu_+ e^{x/l_{\textrm{eq}}}-\nu_-}{\nu_+e^{L/l_{\textrm{eq}}} - \nu_-},
	\end{align}
\end{subequations}
and the ``upstream conductance'' 
\begin{equation}
 \label{eq:Gu}
G_u \equiv  I_{-}(0)/V_0 \simeq 0.
\end{equation}
 Hence, an injected current propagates entirely downstream in the fully equilibrated (incoherent) limit. 
 
\begin{figure}[t]
\includegraphics[width=0.99\columnwidth]{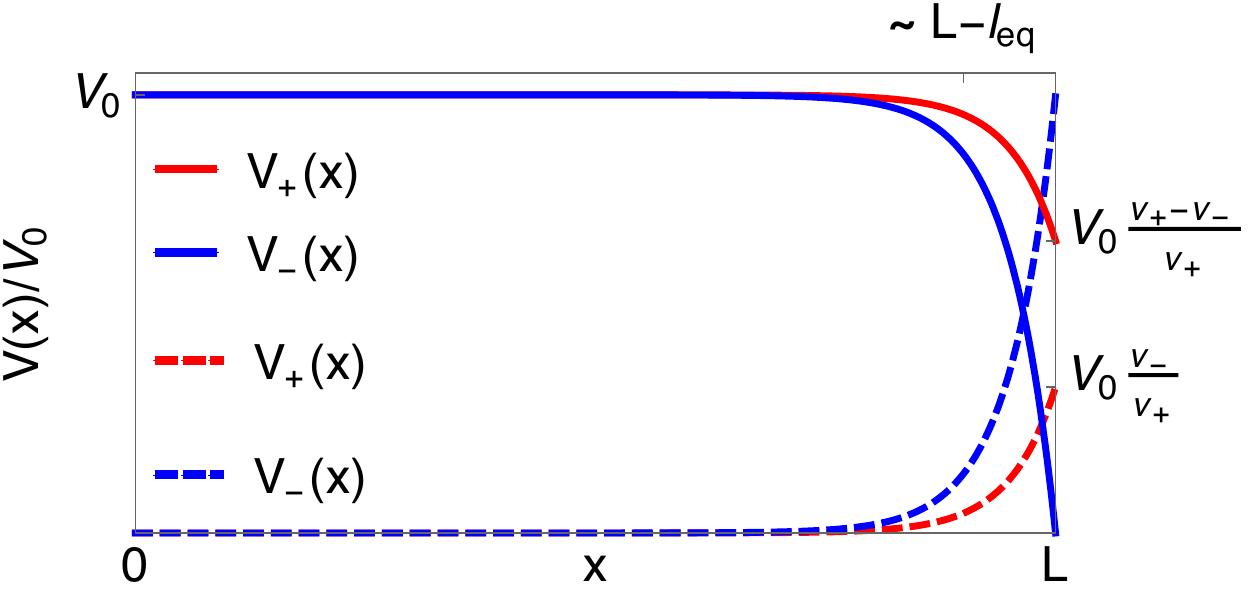}\caption{\label{fig:VoltageProfiles} Voltage profiles for counter-propagating modes $V_+(x)$ (downstream, red lines) and $V_-(x)$ (upstream, blue lines) with filling factor discontinuities $\nu_+$ and $\nu_-<\nu_+$, respectively (see Fig.~\ref{fig:EdgeSegment}). With a voltage bias $V_0$ on the contact at $x=0$ (thick curves), the downstream-mode voltage $V_+(x)$ drops from $V_0$ to $(\nu_+-\nu_-)V_0/\nu_+$ within the length $l_{\rm eq}$ from the right grounded contact ($x=L$) due to equilibration. Biasing instead the contact at $x=L$ (dashed curves) results also in a voltage drop in this region. In this case, the current reaching the other (grounded) contact at $x=0$ is exponentially suppressed in $L/l_{\rm eq}$. For all curves, we have used $L/l_{\rm eq}=20$.}
\end{figure}

 The two-terminal charge conductance $G$ is given by $G = G_d + G_u$. Eqs.~\eqref{eq:Gd}) and~\eqref{eq:Gu} yield (up to an exponentially small correction)
 \begin{equation}
 \label{eq:G-incoherent}
 G = \nu e^2/h,
 \end{equation}
 which is a hallmark of the incoherent regime. We plot the voltage profiles of the edge segment for both choices of boundary conditions in Fig.~\ref{fig:VoltageProfiles}.

By combining Eqs.~\eqref{eq:VProfiles} and~\eqref{eq:VProfiles2}, the fully equilibrated edge segment is described by the following conductance equation
\begin{equation}
\label{eq:LinejunctionConductanceMatrix}
\begin{pmatrix}
I_{+}(L) \\ I_{-}(0) 
\end{pmatrix} = 
\frac{e^2}{h}\begin{pmatrix}
 \nu_{+}-\nu_{-} & \nu_{-} \\
\nu_{-} & 0
\end{pmatrix} \begin{pmatrix}
V_{+}(0) \\ V_{-}(L)  
\end{pmatrix}.
\end{equation}
By using the local current-voltage relations (see Eq.~\eqref{eq:voltage-current})
$V_{+}(0)=I_+(0)h/(\nu_+e^2)$ and $V_{-}(L)= I_-(L)h/(\nu_-e^2)$, we can write an equivalent branching matrix relation in terms of the currents,
\begin{equation}
\label{eq:LinejunctionBranchingMatrix}
\begin{pmatrix}
I_{+}(L) \\ I_{-}(0) 
\end{pmatrix} = 
\begin{pmatrix}
 1-\nu_{-}/\nu_{+} & 1 \\
\nu_{-}/\nu_{+} & 0
\end{pmatrix} \begin{pmatrix}
I_{+}(0) \\ I_{-}(L)  
\end{pmatrix}.
\end{equation}
From Eqs.~\eqref{eq:VProfiles} and~\eqref{eq:VProfiles2} we further note that, for both choices of boundary conditions, the local voltage difference along the edge segment reads
\begin{align} 
\label{eq:Vdiff1}
	\delta V (x)\equiv |V_+(x)-V_-(x)| = \left | V_0\frac{e^{x/l_{\textrm{eq}}}(\nu_+-\nu_-)}{e^{L/l_{\textrm{eq}}}\nu_+-\nu_-} \right |.
\end{align}

From this expression, we conclude that, regardless of the choices of source and drain contacts, the local voltage along the edge drops only within a region of length $\sim l_{\rm eq}$ \textit{in the vicinity of $x=L$} and is therefore fixed by the net chirality, i.e, the downstream direction (see Fig.~\ref{fig:VoltageProfiles}). This voltage drop is associated with Joule heating, and the region is therefore referred to as the \textit{hot spot}~\cite{Tamaki2005,Park2019,Spanslatt2019} (see Fig.~\ref{fig:EdgeSegment}).

The corresponding dissipated power, $P$, can be computed according to
\begin{align}
\label{eq:DissPower}
P &\equiv P_{\rm in}-P_{\rm out} \notag \\
&= \frac{e^2}{2h}(\nu_{+} V_+^2(0)+\nu_{-} V^2_-(L)-\nu_{+} V_+^2(L)-\nu_- V^2_-(0)) \notag \\
	& =\frac{e^2}{h}\frac{(V_+(0)-V_-(L))^2(\nu_{+}-\nu_{-})\nu_{-}}{2 \nu_{+}}. 
\end{align}
As will be shown next,  the nature and the direction of the heat transport away from the hot spot depend crucially on $\nu_Q$ which is fixed by the topological order of the bulk~\cite{Kane1997,Capelli2002}.

\subsection{\label{sec:HeatEquilibration} Heat equilibration}
For the local temperatures, Eqs.~\eqref{eq:ConservationTunneling} and~\eqref{eq:TunnelingCurrents} yield in the continuum limit
\begin{equation}
\label{eq:TemperatureEquation}
\partial_x\vec{T^2}(x) = \mathcal{M}_T \vec{T^2}(x) + \Delta\vec{V}(x),
\end{equation}
where $\vec{T^2}(x)=(T_+^2(x),T_-^2(x))^T$,
\begin{equation}
\label{eq:T2times2}
	\mathcal{M}_{T\;} = \frac{\gamma}{l} \begin{pmatrix}
-\chi_+ n_d & \chi_+ n_d\\
\chi_- n_u & -\chi_- n_u
\end{pmatrix},
\end{equation}
and the Joule heating contribution
\begin{align}
\label{eq:Vdiff2}
\Delta \vec{V}(x)=\frac{e^2 [\delta V(x)]^2 }{h\kappa l}\begin{pmatrix}
	-1 \\
	+1
\end{pmatrix}.
\end{align}
Note that Eq.~\eqref{eq:T2times2} includes $n_d$ and $n_u$ and therefore takes into account the microscopic composition of downstream and upstream propagating channels. It was shown in Ref.~\onlinecite{Spanslatt2019} that Eq.~\eqref{eq:T2times2} allows three different types of heat transport along the edge. 

\begin{enumerate}[i)]
	\item Ballistic heat transport, $n_d>n_u$:
 \begin{equation}
\label{eq:TDecay}
k_B T_{\pm}(x) \sim V_0 \times \left[\mathcal{O}\left(e^{-\frac{\gamma}{2}\frac{\nu_+ \nu_-}{\nu_+-\nu_-}\frac{L-x}{l_{\rm eq}}}\right)+\mathcal{O}\left(e^{-\frac{L-x}{l_{\rm eq}}}\right)\right].
\end{equation}
In this case, the heat propagates ballistically downstream. As a result, the heat propagating upstream from the hot spot is exponentially suppressed in $L/l_{\rm eq}$. The special case $n_{u}=0$ yields exact $T_{\pm}(x) \equiv 0$: without any upstream channels, no heat can propagate upstream.
\item Diffusive heat transport, $n_d=n_u$: 
\begin{equation}
\label{eq:TDiffusive}
k_B T_{\pm}(x) \sim V_0 \times \sqrt{x/L}.
\end{equation}
\item ``Antiballistic'' heat transport, $n_d<n_u$: 
\begin{equation}
\label{eq:TConst}
k_B T_{\pm}(x) \sim V_0 \times {\rm const.}
\end{equation}
In this case, the heat propagates ballistically upstream; we call this type of heat transport ``antiballistic''.
\end{enumerate}

Since the charge transport is always downstream in the incoherent regime, we conclude that there exist exactly three possible combinations of edge transport: i) Both charge and heat flow ballistically downstream; ii) charge flows ballistically downstream and the heat diffuses; iii) charge flows ballistically downstream but heat flows ballistically upstream, i.e. ``antiballistically''. It was shown in Ref.~\onlinecite{Spanslatt2019} that these topologically distinct cases exhibit three different noise characteristics. For completeness, the derivation of this result is outlined next. 

\subsection{\label{sec:Noise Mechanism}Noise generation}
On the FQH edge segment depicted in Fig.~\ref{fig:EdgeSegment}, noise is generated due to partitioning of the electric current by inter-mode charge tunneling (see Refs.~\onlinecite{deJong1996,Blanter2000} for review of shot noise induced by partitioning in various systems). Fluctuations of the local tunneling current of charge can be decomposed into
\begin{equation}
\label{eq:delta-tau}
\delta I^\tau_{j} = \delta I^{\tau,{\rm tr}}_{j} + \delta I^{\tau,{\rm int}}_{j}. 
\end{equation}
Here and below, $\delta X$ denotes the deviation of a quantity $X$ from its time average, $\delta X\equiv X-\overline{X}$. The intrinsic contributions $\delta I^{\tau,{\rm int}}_{j}$  arise from local Johnson-Nyquist noise; we take them to be independent random variables with zero mean and with variance
\begin{equation}
\label{eq:TempFluct}
\overline{\delta I^{\tau,{\rm int}}_{j}\delta I^{\tau,{\rm int}}_{j'}}= \frac{2e^2}{h} g k_B\left(T_{+,j}+T_{-,j'}\right) \delta_{j,j'}.
\end{equation}
We have here assumed that the local voltage difference between the two modes is much smaller than their average temperature: $V_{+,j}-V_{-,j}\ll k_B(T_{+,j}+T_{-,j})/2$. This approximation is excellent (it holds up to exponentially small corrections in $L/l_{\rm eq}$, see Eq.~\eqref{eq:Vdiff1}) around $x=0$, which will be shown below to be the region where most of the noise is generated.
The transmitted contributions  $\delta I^{\tau,{\rm tr}}_{j}$ in Eq.~\eqref{eq:delta-tau} reflect fluctuations in the voltage difference between the channels,
\begin{equation}
\delta I^{\tau,{\rm tr}}_{j} = g \frac{e^2}{h}\left(\delta V_{+,j}-\delta V_{-,j}\right),
\end{equation}
that are induced by $\delta I^{\tau,{\rm int}}_{j}$  according to the transport equation \eqref{eq:VoltageEquation}. 

In the continuum limit, we find the following equation for the local electric current fluctuations
\begin{equation}
\label{eq:NoiseEquation}
\partial_x\vec{\delta I^{}}(x) = \mathcal{M}_I \vec{\delta I^{}}(x) + \vec{\delta I}^{\tau,{\rm int}}(x),
\end{equation}
where 
\begin{equation}
\vec{\delta I}^{\tau,{\rm int}}(x) \equiv \lim_{a \rightarrow 0}  \frac{\delta I_j^{\tau, \rm{int}}}{a} \begin{pmatrix}
	-1 \\
	+1
\end{pmatrix}.
\end{equation}
The noise at the two ends of the edge segments, $S\equiv \overline{ \delta I_+(L)^2}=\overline{ \delta I_-(0)^2}$, becomes~\cite{Park2019}
\begin{equation}
\label{eq:Noise2Modes}
S \simeq \frac{2e^2}{h l_{\rm{eq}}}\frac{ \nu_-}{ \nu_+} (\nu_+-\nu_-) \int_0^L dx\;\frac{e^{-\frac{2x}{l_{\rm{eq}}}}k_B\left(T_+(x) +T_-(x) \right)}{(1-e^{-\frac{L}{l_{\rm{eq}}}}\nu_-/\nu_+)^2},
\end{equation}
where we have used the continuum limit of Eq.~\eqref{eq:TempFluct},
\begin{equation}
\overline{\delta I^{\tau,{\rm int}}(x)\delta I^{\tau,{\rm int}}(y)} = \frac{2e^2g}{h} k_B\left(T_{+}(x)+T_{-}(y)\right)\delta(x-y).
\end{equation}

Equation \eqref{eq:Noise2Modes} is derived under the assumption that the fluctuations in the source contact are negligible (the effect of drain fluctuations can be shown to be exponentially suppressed in $L/l_{\rm eq}$). While the noise in principle originates from the heating along the full edge, Eq.~\eqref{eq:Noise2Modes} shows that the dominant contribution comes from the region of extension  $\sim l_{\rm eq}$ near $x=0$: the \textit{noise spot}. The other contributions  are exponentially suppressed in $x/l_{\rm eq}$. The noise spot is therefore always located on the opposite side of the edge segment with respect to the hot spot (see Fig.~\ref{fig:EdgeSegment}).  

The physical mechanism for the shot noise in the incoherent regime is the following: the voltage drop within $\ell_{\rm{eq}}$ from $x = L$ (the hot spot) produces heat that propagates along the edge. In turn, this heating induces thermally activated tunneling between the edge modes which lead to particle-hole pair excitations. 
Such a pair gives a contribution to the zero-frequency shot noise if its constituents reach different contacts. This happens with a considerable probability if the pair is created within a distance $\sim l_{\rm eq}$ from the $x=0$ contact. Particles and holes created much further away from $x=0$, will eventually flow in the downstream direction (to the $x=L$ contact) in view of the corresponding property of the charge conductances discussed in Sec.~\ref{sec:VoltageEquilibration}. This is reflected in the exponential suppression of contributions from $x \gg  l_{\rm eq}$ to Eq.~\eqref{eq:Noise2Modes}.

Eq.~\eqref{eq:Noise2Modes} also indicates that the characteristics of the noise depends crucially on the temperature profiles $T_\pm(x)$. Substituting the three possible temperature profiles from Sec.~\ref{sec:HeatEquilibration}, we obtain three topologically distinct types of the scaling of the noise in the incoherent regime:

\begin{enumerate}[\itshape i)]
	\item Ballistic heat transport yields exponentially suppressed noise:
	 \begin{equation}
	S\sim (e^3V_0/h) \exp\left(-\frac{\gamma}{2}\frac{\nu_+ \nu_-}{\nu_+-\nu_-}\frac{L}{l_{\rm eq}}\right).
\end{equation}
The special case $n_u=0$ yields identically zero noise. 

\item Diffusive heat transport yields algebraically decaying noise: \begin{equation}
S\sim (e^3V_0/h) \times \sqrt{l_{\rm eq}/L}. 
\end{equation}
\item Antiballistic heat transport yields constant noise: \begin{equation}
S\sim  (e^3V_0/h) \times {\rm const}.
\end{equation}
\end{enumerate}
These distinct noise characteristics were shown in Ref.~\onlinecite{Spanslatt2019} to provide a topological classification of the Abelian FQH states and provides an indirect probe of upstream heat transport on the edge.

\subsection{Key hallmarks of the incoherent regime}
We find it instructive to summarize here the key features of transport through an edge segment in the incoherent regime, $l_{\rm eq}\ll L$:
\begin{itemize}
	\item  The transport coefficients exhibit a robust quantization \eqref{eq:G} and~\eqref{eq:GQ} in the incoherent regime.
	\item The charge propagates ballistically downstream.  The upstream charge transport is suppressed exponentially in $L/l_{\rm eq}$.
	\item The thermal transport is ballistic for $\nu_Q>0$, diffusive for $\nu_Q=0$, and antiballistic for $\nu_Q<0$. In the diffusive ($\nu_Q=0$) case the correction to the quantized (zero) value of the thermal conductance is of order $l_{\rm eq}/ L $.  In the ballistic and antiballistic cases, the correction to the quantized value is exponentially small in $L/l_{\rm eq}$.
	\item The regions where heat and noise are generated are spatially separated since they are located on the opposite ends of the edge segment. The locations of these spots are solely determined by the net chirality (the direction of charge propagation).
	\item The combination of different types of thermal transport with the spatial separation of heat and noise generation leads to three different types of asymptotic shot noise characteristics: $S\simeq 0$ up to exponentially small corrections (for $\nu_Q>0$), $S\sim L^{-1/2}$ ($\nu_Q=0$), or $S\sim {\rm const}$ ($\nu_Q<0$).
\end{itemize}

At this point, the following comment is in order. The transport in the incoherent regime is described in terms of hydrodynamic modes which are solutions to the transport equations~\eqref{eq:VoltageEquation} and~\eqref{eq:TemperatureEquation}. Remarkably, the charge- and heat-carrying degrees of freedom are decoupled and in propagate generally in very different ways. This hydrodynamic charge-energy separation should not be confused with the emergence of heat-carrying ``neutral modes'' in the sense of chargeless eigenmodes to a chiral Luttinger liquid Hamiltonian. A description in terms of such modes requires a {\it coherent renormalization} of the system to an infrared KFP fixed point at which the neutral modes decouple from the charged degrees of freedom~\cite{Kane1994,Kane1995b,Moore1998,Protopopov2017}. As has been discussed above, staying in the coherent regime and reaching the KFP fixed point is a very difficult experimental task since it would require an extremely low temperature. At the same time, the separation into charge and heat hydrodynamic modes in the incoherent regime is ubiquitous in experiments on complex FQH edges. 

In the next two sections, we use our model of the incoherent regime to compute charge conductance and noise in QPC geometries.

\section{Conductance plateaus in QPC geometries}
\label{sec:QPCApplications}

\subsection{\label{sec:Incoherent quantum point contact}Quantum point contact branching matrix}

\begin{figure}[t]
\captionsetup[subfigure]{position=top,justification=raggedright}
\subfloat[]{
\includegraphics[width=0.95\columnwidth]{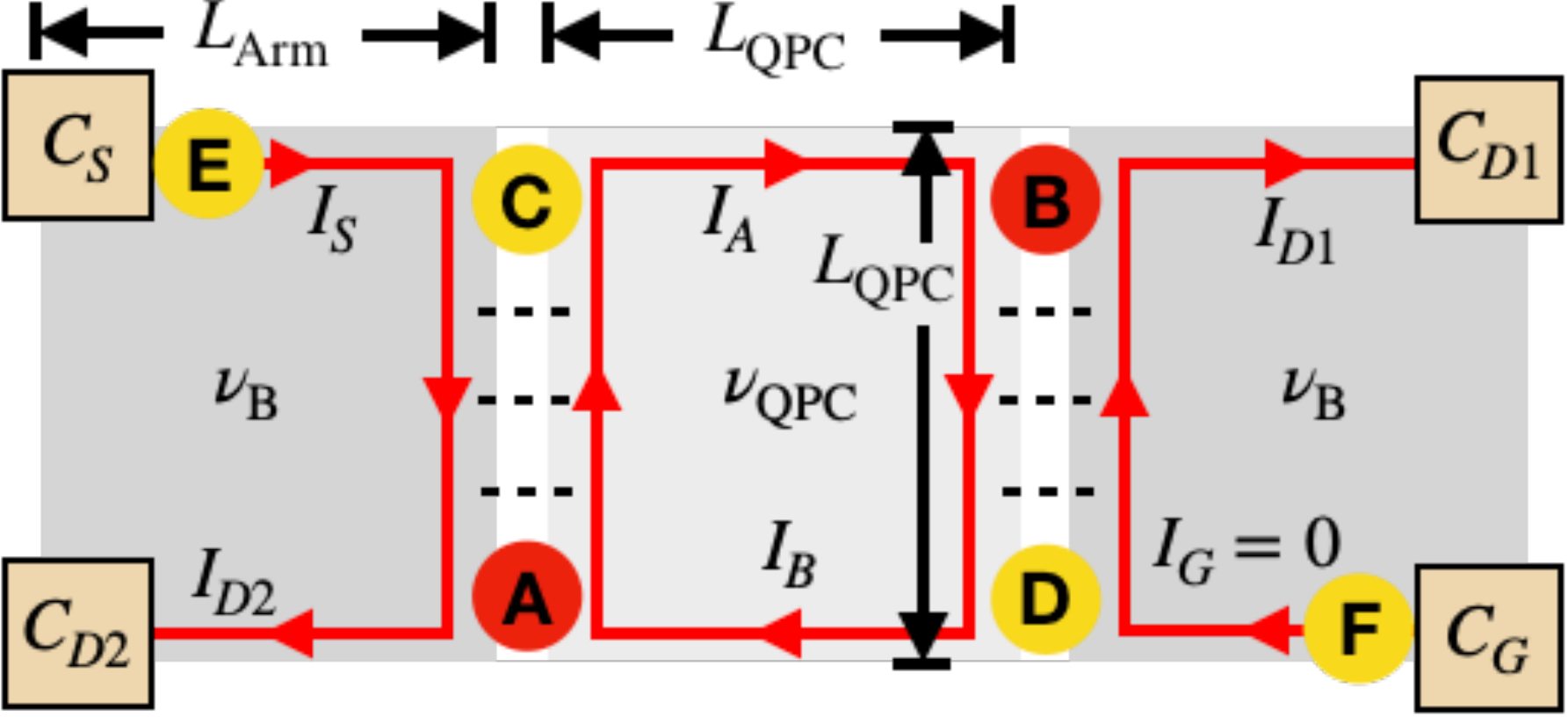}
\label{fig:DQPCa}}
\\[0.0cm]
\subfloat[]{
\includegraphics[width =0.95\columnwidth]{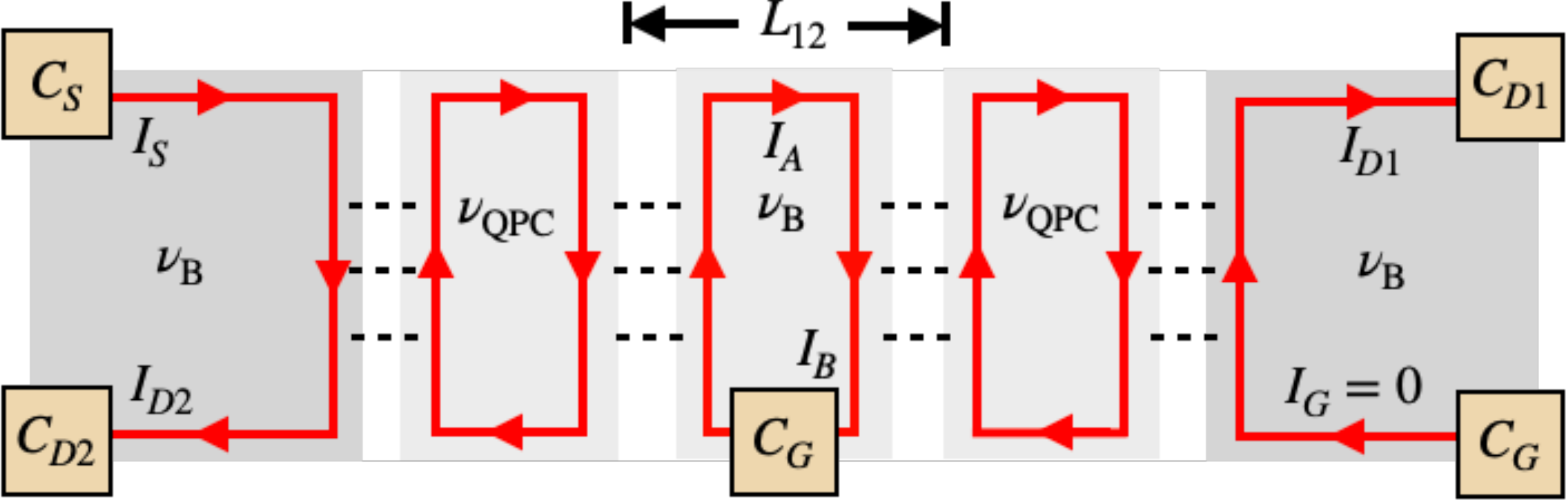}
\label{fig:DQPCb}}
\caption{ (a) Effective geometry with characteristic length scales $L_{\rm Arm}$ and $L_{\rm QPC}$, both assumed to be much longer than the equilibration length $l_{\rm eq}$, for a single QPC or a short-distance double-QPC setup.  The red dots, $A$ and $B$, denote the hot spots where voltage drops occur, while the yellow dots, $C-F$, denote regions where noise may be generated. The conductances and the shot noise are measured at the drain contacts $C_{D1}$ and $C_{D2}$.
 (b) Geometry of a long-distance double QPC. Note the extra grounded contact in the central region. The distance between the low-density regions is $L_{12}$.}
\label{fig:DQPCs}
\end{figure}

We model a single QPC device as a low density constriction with filling factor $\nu_{\rm{QPC}} <\nu_B$ (see Fig.~\ref{fig:DQPCa}) and assume that $l_{eq}\ll L_{\rm{QPC}}\ll L_{\rm Arm}$ so that we can treat this system as fully incoherent (see also Fig.~\ref{fig:LengthScales}). Under this assumption, the QPC is equivalent to two line junctions in series. Within this picture, charge and heat propagate across the device due to successive equilibrations in these line junctions. Upon biasing the source contact $C_S$ with respect to the ground contact $C_G$, the conductances measured in the two drains $C_{D1}$ and $C_{D2}$ can be computed (cf. Ref.~\onlinecite{Milletari2013}) by considering two coupled line junction branching matrix equations~\eqref{eq:LinejunctionBranchingMatrix}: 
\begin{equation}
\label{eq:QPCB1}
\begin{pmatrix}
I_{D2} \\ I_{A} 
\end{pmatrix} = 
\begin{pmatrix}
 1-\nu_{\rm QPC}/\nu_{\rm B} & 1 \\
\nu_{\rm QPC}/\nu_{\rm B} & 0
\end{pmatrix}\begin{pmatrix}
I_{S} \\ I_{B} 
\end{pmatrix}
\end{equation}
and 
\begin{equation}
\label{eq:QPCB2}
\begin{pmatrix}
I_{B} \\ I_{D1} 
\end{pmatrix} = 
\begin{pmatrix}
0 & \nu_{\rm QPC}/\nu_{\rm B} \\
1 & 1-\nu_{\rm QPC}/\nu_{\rm B}
\end{pmatrix}\begin{pmatrix}
I_{A} \\ I_{G} 
\end{pmatrix}.
\end{equation}
Eliminating the internal currents $I_A$ and $I_B$, we obtain the branching matrix equation of the QPC as
\begin{equation}
\label{eq:QPCBranch}
\begin{pmatrix}
I_{D1} \\ I_{D2} 
\end{pmatrix} = 
\begin{pmatrix}
\nu_{\rm QPC}/\nu_{\rm B} & 1-\nu_{\rm QPC}/\nu_{\rm B} \\
1-\nu_{\rm QPC}/\nu_{\rm B} & \nu_{\rm QPC}/\nu_{\rm B}
\end{pmatrix} \begin{pmatrix}
I_{S} \\ I_{G} 
\end{pmatrix}.
\end{equation}
 
The dissipated power in each of the two line junctions can be computed according to Eq.~\eqref{eq:DissPower}. Since the voltage drop only depends on $\nu_{\rm B}$ and $\nu_{\rm QPC}$, both dissipated powers are equal and read
\begin{equation}
\label{eq:PQPC}
	P_{\rm A} = 	P_{\rm B} = \frac{e^2}{h}\frac{V_0^2(\nu_{\rm B}-\nu_{\rm QPC})\nu_{\rm QPC}}{2 \nu_{\rm B}},
\end{equation}
where $V_0$ is the bias voltage in the source contact $C_{S}$. The location of the two hot spots, depicted by the red regions labelled $A$ and $B$ in Fig.~\ref{fig:DQPCa}, are on oppositely oriented corners of the low density region. In principle, there exist two possible additional voltage drop locations (i.e. hot spots) in the setup. Specifically, dissipation takes place also in the drain contacts $C_{D1}$ and $C_{D2}$. However, since the drain contacts are floating with respect to dc currents in the QPC experiments of Refs.~\onlinecite{Bhattacharyya2019} and~\onlinecite{Sabo2017}, those hot spots are neglected in the calculation of noise below; see Appendix~\ref{sec:AppHotspots} for a more detailed discussion of this issue. 

\subsection{\label{sec:SQPC results} Conductance plateaus in single QPC}
To apply our model to the experiment in Ref.~\onlinecite{Bhattacharyya2019}, we begin by setting, without loss of generality, $I_{G}=0$ in Eq.~\eqref{eq:QPCBranch}. This amounts to a choice of ground. We then readily obtain the conductances measured at the drain contacts $C_{D1}$ and $C_{D2}$,
\begin{subequations}
\label{eq:QPCG}
\begin{align}
&G_{D1} = \nu_{\rm QPC}\frac{e^2}{h} \label{eq:GCD1}, \\
&G_{D2} = (\nu_{\rm QPC}-\nu_{\rm B})\frac{e^2}{h}.
\end{align}
\end{subequations}
We see that $G_{D1}$ is independent of $\nu_{\rm B}$. This provides a general explanation for the formation of the experimentally observed $G_{D1} = (1/3)e^2/h$ conductance plateau for a variety of $\nu_{\rm B}$ states. 
Specifically, if for a finite interval of the QPC gate voltage there exists a stable $\nu_{\rm QPC}=1/3$  FQH state in the low density region (which is highly plausible because the $\nu_{\rm QPC}=1/3$ state is the most stable state in the FQH regime), the incoherent transport through the QPC (with full equilibration in the two successive line junctions) leads to $G_{D1} = (1/3)e^2/h$.

Other plateaus emerge within this theory in the same way if stable FQH states corresponding to filling factors $\nu_{\rm QPC}<\nu_{B}$ are formed in the QPC region. The number of observed plateaus and their visibilities will thus strongly depend on the quality of the underlying 2D electron gas and on the temperature. In addition, they may be influenced by the geometry of the low-density region. Indeed, additional plateaus are also observed but they are less prominent than the $G=(1/3)e^2/h$ plateau. This is not surprising since $\nu_{\rm QPC}=1/3$ is the most stable FQH state. 

\subsection{\label{sec:DQPCresults}Conductance plateaus in double QPC}
\begin{figure}[t]
\captionsetup[subfigure]{position=top,justification=raggedright}
\subfloat[]{
\includegraphics[width=0.85\columnwidth]{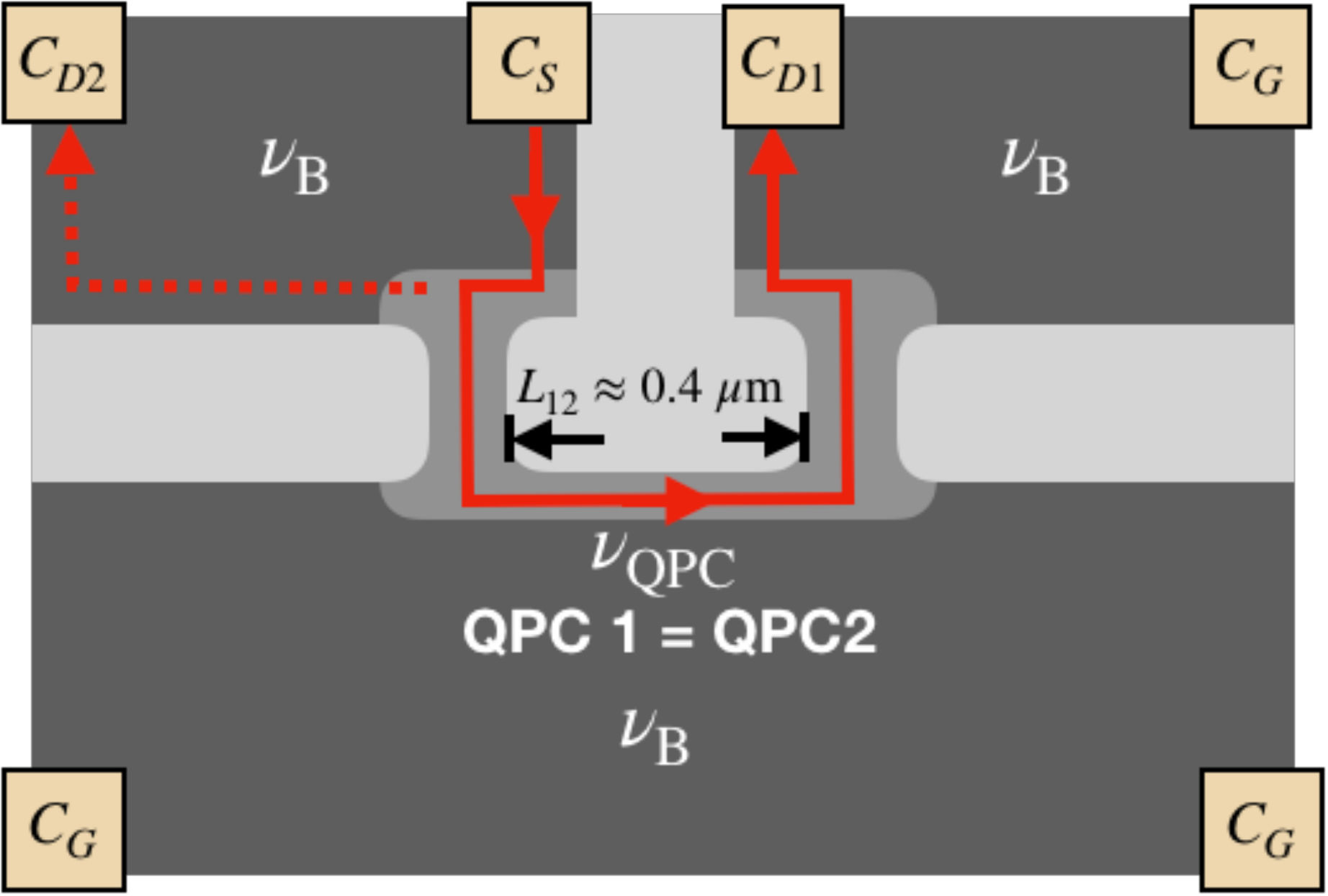}
\label{fig:DQPCExpa}}
\\[0.0cm]
\subfloat[]{
\includegraphics[width =0.85\columnwidth]{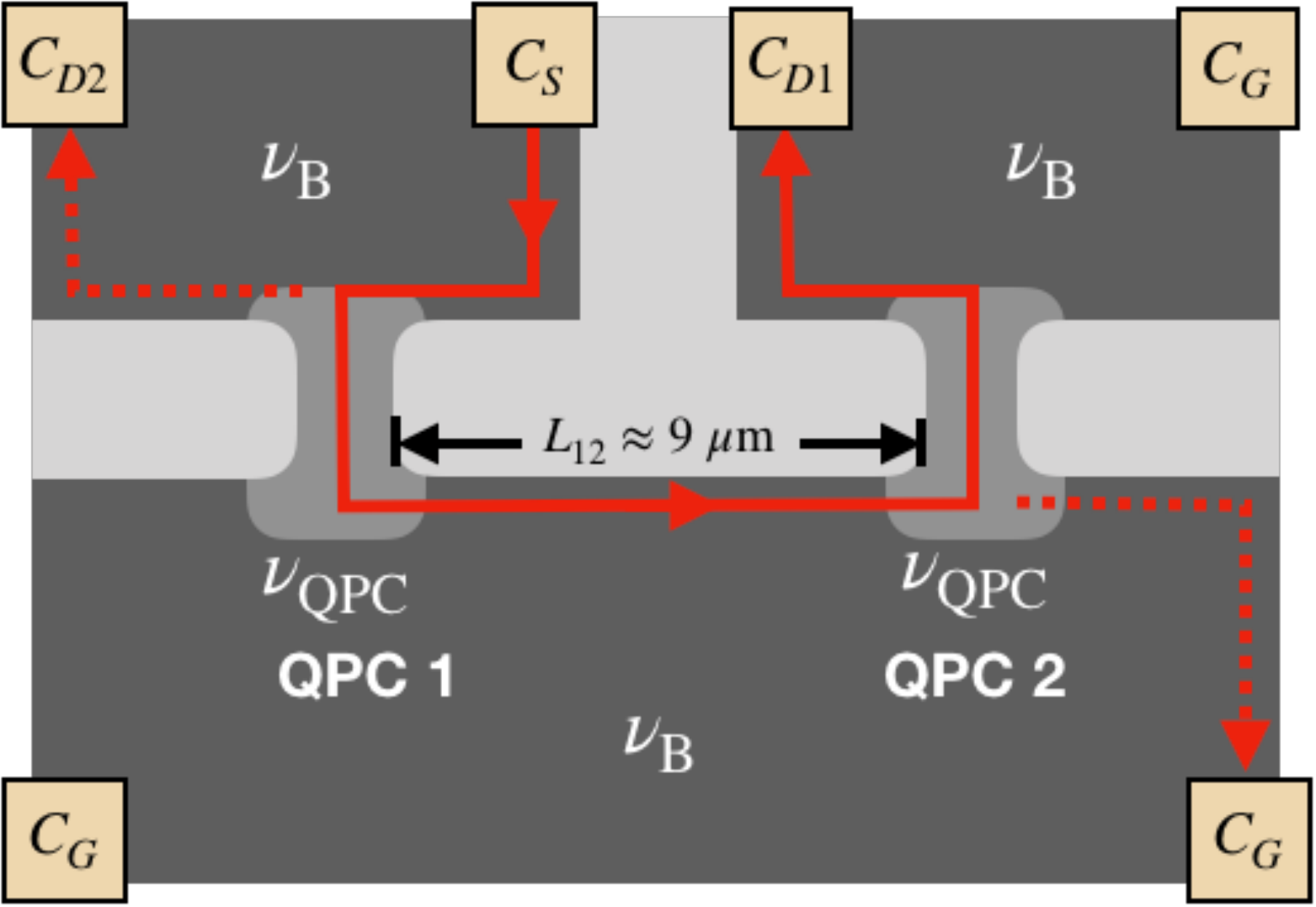}
\label{fig:DQPCExpb}}
\caption{Schematic representation of experimental double-QPC setups in  Ref.~\onlinecite{Sabo2017}. Dark gray regions have filling factors $\nu_{\rm B}$, light gray represent fully depleted regions (i.e. non-FQH regions). Middle-grey regions represent regions depleted to filling factor $\nu_{\rm QPC}<\nu_{\rm B}$. Red thick lines depict the downstream charge current upon injection in the source contact $C_{S}$ and the dashed lines depict the reflected currents at the QPCs. (a) When the distance between the QPCs, $L_{12}$, is sufficiently short, a single low-density ($\nu_{\rm QPC}$) region is formed and the setup is similar to the single QPC in Fig.~\ref{fig:DQPCa}.  (b) Setup with a long distance $L_{12}$ between the QPCs gives rise to two separated low-density regions as modelled in Fig.~\ref{fig:DQPCb}.}
\label{fig:DQPCExp}
\end{figure}

We next apply our model to the double-QPC setup in Ref.~\onlinecite{Sabo2017} (see Fig.~\ref{fig:DQPCExp}). In our formalism, such a device can be treated as two identical QPCs in series. We will assume that both QPC have identical properties, inducing a depletion region with the same filling factor $\nu_{\rm QPC}$.
 In line with the experiment where the distance between the QPCs was varied, we distinguish between two different scenarios: i) The distance between the two low-density regions  is sufficiently small such that they constitute a single region (see Fig.~\ref{fig:DQPCExpa}). The effective model then becomes identical to that of a single QPC depicted in Fig.~\ref{fig:DQPCa}.
 ii)  The distance between the two low-density regions is sufficiently large, so that they are separated by a region with filling factor $\nu_{\rm B}$ (see Fig.~\ref{fig:DQPCExpb}). 
The effective model for this case is shown in Fig.~\ref{fig:DQPCb}.
 The experimental setup is operated with a grounded contact on the edge of the central region (implemented in Fig.~\ref{fig:DQPCb} by the central ground contact $C_{G}$). 

In the first scenario (two closely located QPC's), we can straight away use the QPC branching matrix~\eqref{eq:QPCBranch}, leading to conductances
\begin{align} 
& G_{D1, \rm{short}}=\nu_{\rm QPC}\frac{e^2}{h},    \label{eq:G1-two-QPC-short} \\
& G_{D2, \rm{short}}=(\nu_{\rm B}-\nu_{\rm QPC})\frac{e^2}{h}.   \label{eq:G2-two-QPC-short}
\end{align}
In the second scenario (two distant QPCs), we obtain the two coupled matrix equations
\begin{equation}
\label{eq:DQPC1}
\begin{pmatrix}
I_{A} \\ I_{D2} 
\end{pmatrix} = 
\begin{pmatrix}
\nu_{\rm QPC}/\nu_{\rm B} & 1-\nu_{\rm QPC}/\nu_{\rm B} \\
1-\nu_{\rm QPC}/\nu_{\rm B} & \nu_{\rm QPC}/\nu_{\rm B}
\end{pmatrix} \begin{pmatrix}
I_{S} \\ 0 
\end{pmatrix}
\end{equation}
and
\begin{equation}
\label{eq:DQPC2}
\begin{pmatrix}
I_{B}\\I_{D1} 
\end{pmatrix} = 
\begin{pmatrix}
\nu_{\rm QPC}/\nu_{\rm B} & 1-\nu_{\rm QPC}/\nu_{\rm B} \\
1-\nu_{\rm QPC}/\nu_{\rm B} & \nu_{\rm QPC}/\nu_{\rm B}
\end{pmatrix} \begin{pmatrix}
0 \\I_{A} 
\end{pmatrix},
\end{equation}
where we have taken into account the additional central ground contact. Eliminating the internal-region currents $I_A$ and $I_B$,  we find that the conductances measured at the drain contacts read
\begin{subequations}
	\begin{align}
		& G_{D1, \rm{long}}=\frac{\nu_{\rm QPC}^2}{\nu_{\rm B}}\frac{e^2}{h},  \label{eq:G1-two-QPC-long} \\
		& G_{D2, \rm{long}}=(\nu_{\rm B}-\nu_{\rm QPC})\frac{e^2}{h}.  \label{eq:G2-two-QPC-long}
	\end{align}
\end{subequations}
We note that for the single or short length double QPC, the injected current $I_{S}$ is split into the two drains $C_{D1}$ and $C_{D2}$ (i.e. $I_B=0$ for $I_G=0$). For the long distance double QPC, on the other hand, the grounded contact in the central region (see Fig.~\ref{fig:DQPCb}) also collects a fraction of the injected current: 
\begin{equation}
I_{B}=e^2V_0(\nu_{\rm QPC}-\nu_{\rm QPC}^2/\nu_{\rm B})/h.
\end{equation}
We now consider some examples, which allow us to compare with the experiment of Ref.~\onlinecite{Sabo2017}. Let us assume that each of the two QPCs separately is at the $G = (1/3) e^2/h$ plateau, i.e., $\nu_{\rm QPC}=1/3$.  In Fig. 2 of Ref.~\onlinecite{Sabo2017}, yellow curves in all panels correspond to this situation.
Let us first take the bulk filling factor to be $\nu_{\rm B}=2/3$ as in Fig. 2a,b of Ref.~\onlinecite{Sabo2017}. For the case of two closely located QPCs we find then according to Eq.~\eqref{eq:G1-two-QPC-short} the double-QPC conductance $G_{D1, \rm{short}}=(1/3) e^2/h$, in full agreement with Fig. 2b of Ref.~\onlinecite{Sabo2017} (according to the notation of Ref.~\onlinecite{Sabo2017}, one should multiply the transmission shown in their Fig. 2 by the filling factor $\nu_{\rm B}$ in order to get the conductance $G_{D1}$).
For the case of two QPCs separated by a large distance, we get from Eq.~\eqref{eq:G1-two-QPC-long} the conductance $G_{D1, \rm{long}}=(1/6)e^2/h$, in full agreement with
the plateau on the yellow curve in Fig. 2a of Ref.~\onlinecite{Sabo2017}. 

We further take the bulk filling factor to be $\nu_{\rm B}=3/5$ as in Fig. 2c,d of Ref.~\onlinecite{Sabo2017}. The value of $G_{D1, \rm{short}}$ in our theory remains $(1/3)e^2/h$, which is in perfect agreement with Fig. 2d of Ref.~\onlinecite{Sabo2017}.  For the case of two distant QPCs, we get $G_{D1, \rm{long}}=(5/27)e^2/h$, which is again in a very good agreement with the plateau on the yellow line in Fig. 2c of Ref.~\onlinecite{Sabo2017}.

Finally, we consider the case when the current is injected in the integer quantum Hall regime, $\nu_{\rm B}=1$, still keeping a symmetric double QPC with $\nu_{\rm QPC}=1/3$. In this situation our results predict $G_{D1, \rm{short}}=(1/3)e^2/h$ and $G_{D1, \rm{long}}=(1/9)e^2/h$. We hope that these predictions can be tested in future experiments. 

\section{Noise in single-QPC geometry}
\label{sec:NoiseQPCs}
In this section, we study generation of shot noise in the single QPC geometry outlined in Sec.~\ref{sec:QPCApplications} (see Fig.~\ref{fig:DQPCa}). We will show that the characteristics of the noise are determined by the nature of the  thermal transport (controlled by the signs of the topological thermal coefficients $\nu_{Q}$ in Eq.~\eqref{eq:GQ}) in the outer arms (consisting of the edge channels between the contacts and the QPC), the line junctions (i.e. the interfaces between the low density region and the bulk regions), and the upper and lower edges in the QPC. Hence, the noise can be classified according to possible combinations of heat transport along these edge segments (see Tab.~\ref{tab:QPCQualitative}).

We also present below a general expression for the noise and apply it to the specific configurations $(\nu_{\rm B},\nu_{\rm QPC}) = (3/5,1/3)$ and $(\nu_{\rm B},\nu_{\rm QPC}) = (4/7,1/3)$ for comparison with the experiments in Ref.~\onlinecite{Bhattacharyya2019}. Two other configurations are treated in Appendix~\ref{sec:AppNoise}.

\subsection{Topological noise classification}
\label{sec:QuantNoise}
\begin{table*}[t] 
\caption{\label{tab:QPCQualitative}%
Asymptotics of the normalized noise $S/(e^3 V_0/h)$ (where $S$ is the shot noise, and $V_0$ is the bias voltage) in the  incoherent regime, 
$l_{\rm{eq}} \ll L_{\rm{QPC}} \ll L_{\rm{Arm}}$,
for the possible combinations of the heat transport (B = Ballistic, D = Diffusive, AB = Antiballistic) along the outer arms, the line junctions between the QPC region and the bulk, and the upper and lower edges of the low density region. Ballistic (antiballistic) heat transport is defined as heat propagating in the same (opposite) direction with respect to the charge transport. 
}
\begin{ruledtabular}
\begin{tabular}{lllll}
Outer arms & Line junctions & Central upper/lower edges & Noise asymptotics & Possible realizations
\\ \hline
\multirow{5}{*}{B}  & \multirow{3}{*}{B} & B & 0 &  $\nu_B = 2/5$, $\nu_{\rm{QPC}} = 1/3$  \\ 
&  & D &  $\sqrt{\frac{l_{\rm eq}}{L_{\rm QPC}}}$&  $\nu_B = 1$, $\nu_{\rm QPC} = 2/3$
\\ 
&  & AB & const.  & $\nu_B = 1$, $\nu_{\rm{QPC}} = 3/5$   \\ \cline{2-5}
& D & B  & $\sqrt{\frac{l_{\rm eq}}{L_{\rm QPC}}}$ & $\nu_B = 1$, $\nu_{\rm{QPC}} = 1/3$ \\ 
& AB& B& const. & $\nu_B = 1$, $\nu_{\rm QPC} = 2/5$ \\  \hline
\multirow{3}{*}{D}& B& AB& $\sqrt{\frac{L_{\rm Arm}}{l_{\rm eq}}}$& $\nu_B = 2/3$, $\nu_{\rm QPC} = 3/5$ \\ 
& D& D& $\sqrt{\frac{L_{\rm Arm}}{L_{\rm QPC}}}$& $\nu_B = 4/5$, $\nu_{\rm QPC} = 2/3$ \\ 
& AB & B& $\sqrt{\frac{L_{\rm Arm}}{l_{\rm eq}}}$&  $\nu_B = 2/3$, $\nu_{\rm QPC} = 1/3$\\ \hline
\multirow{5}{*}{AB}& B & AB& const. & $\nu_B = 3/5$, $\nu_{\rm QPC} = 4/7$ \\ 
& D& AB & const. + $ \sqrt{\frac{l_{\rm{eq}}}{L_{\rm{QPC}}}}$ & $\nu_B =11/7$, $\nu_{\rm QPC} = 3/5$  \\ \cline{2-5}
& \multirow{3}{*}{AB} & B & const. & $\nu_B = 3/5$, $\nu_{\rm{QPC}} = 1/3$ \\ 
& &D&const. + $ \sqrt{\frac{l_{\rm{eq}}}{L_{\rm{QPC}}}}$ &$\nu_B = 11/7$, $\nu_{\rm{QPC}} = 2/3$ \\ 
&& AB & const. & $\nu_B = 4/7$, $\nu_{\rm{QPC}} =3/11 $ \\ 
\end{tabular}
\end{ruledtabular}
\end{table*}

In the incoherent regime, noise in the QPC is generated by charge partitioning from thermally activated tunneling at four different noise spots (denoted by the yellow circles $C$, $D$, $E$, and $F$ in Fig.~\ref{fig:QPCNoise}). A tunneling event yields a contribution to the noise if the respective constituents of a particle-hole pair generated by inter-channel tunneling reach different drains. In analogy with the discussion of the noise in a single segment of the edge, this happens with a sizeable probability in the vicinity of noise spots where particles and holes may travel in different directions and eventually end up in different drains $C_{D1}$ and $C_{D2}$.  On the other hand, a pair partitioned far from the noise spot eventually propagates in the downstream direction, reaches the same contact, and thus does not contribute to the (dc) noise.

The amount of generated noise crucially depends on the nature of the heat reaching the noise spots as specified in Sec.~\ref{sec:Noise Mechanism}. When all source contacts have the same base temperature, heating in the device is generated solely due to voltage drops, which only occur at the hot spots due to the chiral nature of the charge transport. The hot spots in the QPC geometry are depicted as the red dots $A$ and $B$ in Fig.~\ref{fig:QPCNoise}. The generated heat may propagate to the noise spots and in turn generate the noise that is measured in $C_{D1}$ and $C_{D2}$. How much heat is transported to the noise spots is dictated by the signs of the thermal coefficients $\nu_{Q, B}$ in the outer arms, $\nu_{Q, \rm{QPC}}$ in the upper and lower edges in the QPC, and $(\nu_{Q, B} - \nu_{Q, \rm{QPC}})$ in the line junctions. 

Each of the thermal coefficients $\nu_{Q, B}$, $\nu_{Q, \rm{QPC}}$, and $(\nu_{Q, B} - \nu_{Q, \rm{QPC}})$ can be either positive, or negative, or zero. Taking into account that 
$\nu_{Q, B}$ is the sum of the other two coefficients, we find 13 possible combinations that are listed in Table~\ref{tab:QPCQualitative}. We use this to classify the noise asymptotics in the incoherent regime $l_{\rm{eq}}\ll L_{\rm QPC}\ll L_{\rm Arm}$  (see Fig.~\ref{fig:IntroFig} for the definition of the lengths involved). We use the same terminology as introduced  in Sec.~\ref{sec:HeatEquilibration}: ballistic (antiballistic) heat flow  indicates that the heat transport is downstream (resp. upstream), i.e. along (resp. opposite to) the direction of the charge transport. We use abbreviations ``B'', ``AB'', and ``D'' for the ballistic, antiballistic, and diffusive heat transport, respectively. 

We begin by considering QPC configurations supporting ballistic heat transport along all edge segments, which is the (B, B, B) line of Table~\ref{tab:QPCQualitative}. In this case, the heat generated at the hot spots flows downstream directly to the contacts $C_{D1}$ and $C_{D2}$, and does not reach the noise spots since they are located upstream from the hot spots. Only an amount of heat exponentially suppressed in $L_{\rm{QPC}}/l_{\rm{eq}}$ reaches the noise spots, which leads to vanishing noise up to exponential corrections. Examples for this fully ballistic scenario include any of the bulk filling fractions $\nu_{\rm{B}}=2/5$, $\nu_{\rm{B}}=3/7$, or $\nu_{\rm{B}}=4/9$, in combination with $\nu_{\rm QPC}=1/3$. Indeed, no noise was detected on the $G = (1/3) e^2/h$ plateau (implying $\nu_{\rm QPC}=1/3$ in our model) for these filling fractions~\cite{Bhattacharyya2019}. 

We next consider QPC configurations with diffusive heat transport on the outer arms while the other segments have either ballistic or antiballistic heat transport. 
There are two such lines in Tab.~\ref{tab:QPCQualitative}:  (D, B, AB) and (D, AB, B). An example for such a situation is the combination $(\nu_{\rm{B}}$, $\nu_{\rm{QPC}}) = (2/3, 1/3)$; the corresponding pattern of the heat flow is shown in Fig.~\ref{fig:Heat23}. 
While the applied bias voltage gives rise to steady heating at the hot spots, the escape of the generated heat  from the QPC region to the contacts is very slow due to the diffusive nature of heat transport in the outer arms.  This slow escape is dictated by a small  heat conductance $\sim l_{\rm{eq}}/L_{\rm{Arm}} \ll 1$ of the outer arm. 
 By solving the heat equations in a self-consistent way~(see Appendix~\ref{sec:AppNoise} for more details),  
we find that the steady state temperatures at the hot spots are proportional to  $\sqrt{L_{\rm{Arm}}/l_{\rm{eq}}}$. The heat generated at the hot spot flows via the AB segments to the noise spots $C$ and $D$ that thus also acquire a temperature  proportional to  $\sqrt{L_{\rm{Arm}}/l_{\rm{eq}}}$. 
 The noise $S$ then becomes
\begin{align}
\label{eq:diffusivenoise1}
S \propto (T_C + T_D ) \propto \sqrt{\frac{L_{\rm{Arm}}}{l_{\rm{eq}}}}.
\end{align}

We consider now the case when the heat transport is diffusive in all the segments, which is the line (D, D, D) of Table~\ref{tab:QPCQualitative}. 
The enhancement of the temperature of the hot spots is the same as in the previous case. At the same time,  the temperatures at the noise spots $C$ and $D$  is now suppressed by a factor $\sim \sqrt{l_{\rm{eq}}/L_{\rm{Arm}}}$ as compared with those of the hot spots. Such a configuration thus leads to the noise asymptotics
\begin{equation} 
\label{eq:diffusivenoise2}
	S \propto (T_C + T_D) \propto  \sqrt{\frac{L_{\rm{Arm}}}{l_{\rm{eq}}}} \sqrt{\frac{l_{\rm{eq}}}{L_{\rm{QPC}}}} = \sqrt{\frac{L_{\rm{Arm}}}{L_{\rm{QPC}}}}.
\end{equation}

It follows from  Eqs.~\eqref{eq:diffusivenoise1} and \eqref{eq:diffusivenoise2} that in the case of diffusive heat transport in the outer arms, the effective Fano factor (as defined by Eq.~\eqref{eq:FanoFactor}) is super-Poissonian Fano factors, $F \gg 1$.  This strong noise is related to slowness of the leakage of the heat generated at the QPC to the external contacts in such configurations. 

We now move on to treat QPC configurations with ballistic  heat transport along the outer arms but  diffusive transport in the line junctions or in the upper/lower edges in the QPC region. There are two such lines in the Table~\ref{tab:QPCQualitative}:  (B, B, D) and (B, D, B). One example of such a configuration is $(\nu_{\rm{B}}, \nu_{\rm{QPC}}) = (1, 1/3)$; the corresponding pattern of the heat flow is shown in Fig.~\ref{fig:Heat1}.
Since the generated heat directly can flow out of the QPC region, the hot spot temperatures in the steady state are found to be constant in the sense that they do not depend on any length scales of the system. Due to the diffusive line junctions, the temperatures at the noise spots $C$ and $D$ are suppressed by a factor $\sim \sqrt{l_{\rm{eq}}/L_{\rm{QPC}}} \ll 1$ compared with those of the hot spots. Thus, the generated noise asymptotics becomes
\begin{align}
S \propto (T_C + T_D ) \propto \sqrt{\frac{l_{\rm{eq}}}{L_{\rm{QPC}}}}.
\label{eq:S_BBD}
\end{align}

Now we turn to the case when all heat flow types are ballistic or antiballistic, with at least one of the segments showing the antiballistic transport. This includes the following five lines of  Table~\ref{tab:QPCQualitative}: (B, B, AB), (B, AB, B), (AB, B, AB), (AB, AB, B), and (AB, AB, AB). An example of such a configuration is  $(\nu_{\rm{B}}, \nu_{\rm{QPC}}) = (3/5, 1/3)$; the corresponding heat-flow pattern is depicted in Fig.~\ref{fig:Heat35}. The hot-spot temperatures are constant (i.e, length-independent) as in the previous case. Further, the heat propagates efficiently from a hot spot to at least one of the noise spots via an AB segment, so that the temperature at the noise spots, and thus the generated heat, is constant as well. In the case that the heat flows antiballistically on the outer arms, heat also arrives at the noise spots $E$ and $F$, yielding an additional constant contribution to noise. Since only ballistic and antiballistic transport of heat is involved in this class of configurations, corrections to the constant noise are exponentially small. A quantitative calculation of the noise for $(\nu_{\rm{B}}, \nu_{\rm{QPC}}) = (3/5, 1/3)$  and $(4/7, 1/3)$ is presented in
Sec.~\ref{sec:noise_subsectionC}.

Finally, we consider the remaining two lines of  Table~\ref{tab:QPCQualitative}: (AB, D, AB) and (AB, AB, D).  The only difference in comparison with the preceding case is that the heat can propagate from the hot spot to the noise spots not only along the AB segments but also (in parallel) along the diffusive ones. As a result, a power-law correction $\sim \sqrt{l_{\rm{eq}}/L_{\rm{QPC}}}$ to a constant noise emerges.

\subsection{General expression for the noise}
\label{sec:QualNoise}
We derive now a general expression for the noise measured on the conductance plateaus in contacts $C_{D1}$ and $C_{D2}$. This expression allows us to present quantitative values for the noise characterized in Sec.~\ref{sec:QuantNoise}. 

\begin{figure}[t]
\includegraphics[width=0.95\columnwidth] {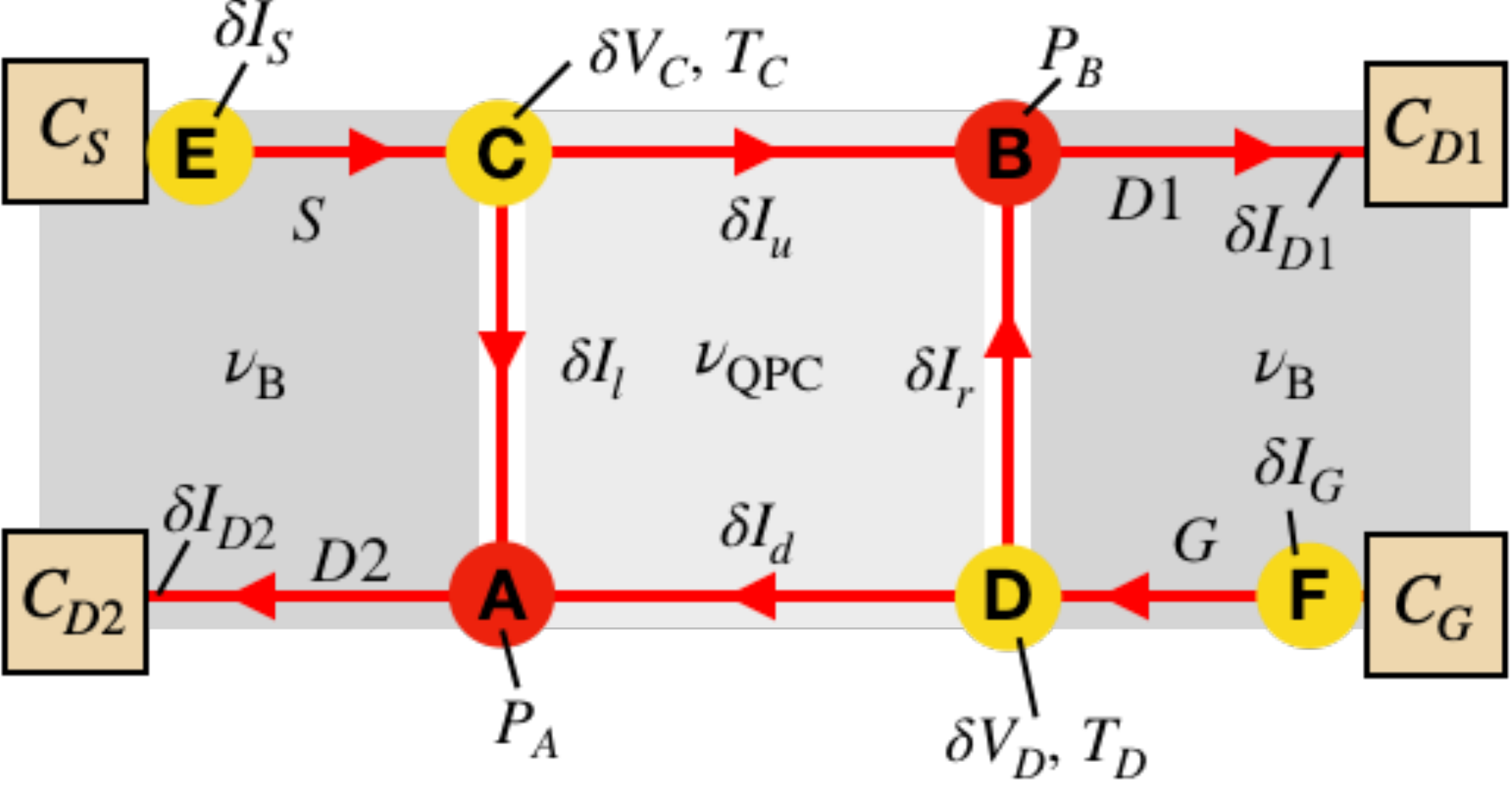}\caption{\label{fig:QPCNoiseAB} Current fluctuations in the incoherent QPC geometry. The chiralities of the charge propagation are depicted by red arrows. $\delta I_{S}$ and $\delta I_{G}$ denote the current fluctuations generated at the noise spots $E$ and $F$ (cf. Fig.~\ref{fig:QPCNoise}), $\delta I_{u,d,l,r}$ denote current fluctuations in the central region, and $\delta V_{C}$ and $\delta V_{D}$ denote fluctuations in the electrochemical potentials at the noise spots $C$ and $D$ respectively. The temperatures at the noise spots are $T_C$ and $T_{D}$ and the dissipated power at the hot spots $A$ and $B$ are $P_A$ and $P_B$, respectively.}
\label{fig:QPCNoise}
\end{figure}

The current fluctuations generated at the noise spots $E$ and $F$ are denoted as $\delta I_{S}$ and $\delta I_{G}$, respectively (see Fig.~\ref{fig:QPCNoiseAB}).
Current conservation at the noise spots $C$ and $D$ yields
\begin{subequations}	
\label{eq:currentIS}
\begin{align}
\delta I_S & = \delta I_{\rm{u}} + \delta I_{\rm{l}}, \\ \label{eq:currentIG}
\delta I_G & = \delta I_{\rm{d}} + \delta I_{\rm{r}},
\end{align}
\end{subequations}
where $\delta I_{\rm{u/d/l/r}}$ are the fluctuations in the currents exiting the noise spots on the up, down, left, and right segments of the QPC region (see Fig.~\ref{fig:QPCNoiseAB}).
Each of these fluctuations can in turn be decomposed into two contributions:
\begin{subequations}
\label{eq:fluctcomp}	
\begin{align}
\delta I_{\rm{l}} &= (\nu_{B} - \nu_{\rm{QPC}}) \frac{e^2}{h} \delta V_C + \delta I^{\rm{th}}_{\rm{l}},\\
\delta I_{\rm{u}}& = \nu_{\rm{QPC}} \frac{e^2}{h} \delta V_C + \delta I^{\rm{th}}_{\rm{u}},\\
\delta I_{\rm{r}} &= (\nu_{B} - \nu_{\rm{QPC}}) \frac{e^2}{h} \delta V_D
 + \delta I^{\rm{th}}_{\rm{r}}, \\
\delta I_{\rm{d}} & = \nu_{\rm{QPC}} \frac{e^2}{h} \delta V_D + \delta I^{\rm{th}}_{\rm{d}}.
\end{align} 
\end{subequations}
The first contribution follows from the fluctuations of the effective electrochemical potentials $\delta V_C$ and $\delta V_{D}$ at the respective noise spots, and the second, denoted as $\delta I_{\rm{l,u,r,d}}^{\rm{th}}$, comes from  non-zero temperature fluctuations. Employing Eqs.~\eqref{eq:currentIS}-\eqref{eq:fluctcomp}, we can write $\delta V_C$ and $\delta V_{D}$ as
\begin{align} \label{eq:voltageC}
\nu_{B}\frac{e^2}{h} \delta V_C = \delta I_S - \delta I^{\rm{th}}_{\rm{l}}- \delta I^{\rm{th}}_{\rm{u}}, \\ \label{eq:voltageD}
\nu_{B}\frac{e^2}{h} \delta V_D= \delta I_G - \delta I^{\rm{th}}_{\rm{r}}- \delta I^{\rm{th}}_{\rm{d}}.
\end{align}

The current fluctuations at the contact $C_{D1}$ read
\begin{align} \label{eq:Currentfluc_D1}
\delta I_{D1} &= \delta I_{\rm{u}} + \delta I_{\rm{r}} \nonumber \\ 
&=  \left [ \nu_{\rm{QPC}} \frac{e^2}{h} \delta V_C + \delta I_{\rm{u}}^{\rm{th}}\right ] \nonumber \\ &
 + \left [(\nu_{B} - \nu_{\rm{QPC}}) \frac{e^2}{h} \delta V_D + \delta I_{\rm{r}}^{\rm{th}} \right ].
\end{align}
Inserting Eqs.~\eqref{eq:voltageC} and \eqref{eq:voltageD} into Eq.~\eqref{eq:Currentfluc_D1},
we obtain $\delta I_{D1}$ given in terms of the independent current fluctuations:
\begin{align}
\delta I_{D1} &= \frac{\nu_{\rm{QPC}}}{\nu_{B}} \delta I_S + 
\frac{\nu_{B}-\nu_{\rm{QPC}}}{\nu_{B}} \delta I_{G} \nonumber \\ 
& + \frac{\nu_{B}-\nu_{\rm{QPC}}}{\nu_{B}} (\delta I_{\rm{u}}^{\rm{th}} - \delta I_{\rm{d}}^{\rm{th}}) + \frac{\nu_{\rm{QPC}}}{\nu_{B}}  (\delta I_{\rm{r}}^{\rm{th}} - \delta I_{\rm{l}}^{\rm{th}}).
\end{align}
The zero frequency shot noise  $S_{D1}\equiv \overline{(\delta I_{D1})^2}$ is then given by
\begin{align} \label{eq:noiseD1}
&S_{D1} = S_{D1, \rm{QPC}} + S_{D1, \rm{Contact}}, \nonumber \\  
&S_{D1, \rm{QPC}}  = \frac{2e^2}{h} \frac{\nu_{\rm{QPC}}}{\nu_{B}} 
(\nu_{B}-\nu_{\rm{QPC}}) k_B (T_C + T_D), \nonumber \\
&S_{D1, \rm{Contact}}  =  \frac{1}{\nu_{B}^2}
\left (\nu_{\rm{QPC}}^2 \overline{(\delta I_S)^2}+ (\nu_{B} - \nu_{\rm{QPC}})^2
\overline{(\delta I_G)^2} \right).
\end{align}
As is clear from Eq.~(\ref{eq:noiseD1}), the noise $S_{D1}$ is given by a sum of two contributions:
$S_{D1, \rm{QPC}}$, which is generated in the noise spots $C$ and $D$,
and $S_{D1, \rm{Contact}}$, which is generated in $E$ and $F$. In the derivation of Eq.~\eqref{eq:noiseD1}, we have used the local Johnson-Nyquist noise relations 
\begin{eqnarray*}
\overline{(\delta I_{\rm{u}}^{\rm{th}})^2} &=& 2 e^2 \nu_{\rm{QPC}} k_B T_{C} /h, \\
\overline{(\delta I_{\rm{l}}^{\rm{th}})^2} &=& 2 e^2 (\nu_{B} - \nu_{\rm{QPC}}) k_B T_{C} /h, \\
\overline{(\delta I_{\rm{d}}^{\rm{th}})^2} &=& 2 e^2 \nu_{\rm{QPC}} k_B T_{D} /h,  \\
\overline{(\delta I_{\rm{r}}^{\rm{th}})^2} &=& 2 e^2 (\nu_{B} - \nu_{\rm{QPC}}) k_B T_{D} /h, 
\end{eqnarray*}
and the fact that all mutual correlations between these thermal noises are uncorrelated. The zero frequency noise $S_{D2}$ measured at the contact $C_{D2}$ is identical to $S_{D1}$, Eq.~(\ref{eq:noiseD1}), in view of current conservation, as can be also checked by a direct calculation.

The Fano-factor $F$ for the noise at the contact $C_{D1}$ ($C_{D2}$) is defined by the relation
\begin{equation} \label{eq:Fanofactor}
S_{D1} = S_{D2} = 2FeI_{\rm{imp}}\tau(1-\tau),
\end{equation}
where $I_{\rm{imp}}$ is the current impinging on the QPC, i.e. $I_{\rm{imp}}= e^2\nu_{B} V_0/h$. The transmission $\tau\equiv I_{D1}/I_{\rm{imp}}=\nu_{\rm QPC}/\nu_{B}$. Inserting these expressions to Eq.~\eqref{eq:Fanofactor}, we finally obtain the Fano factor expressed in terms of the noise as
\begin{align}
\label{eq:FanofactorQPC}
& F \equiv F_{\rm{QPC}} + F_{\rm{Contact}},\notag\\
& F_{\rm{QPC}} = \frac{S_{D1, \rm{QPC}} \ \nu_{B} h}{2e^3 V_{0}\nu_{\rm QPC}(\nu_{B}-\nu_{\rm QPC})}, \notag\\ 
& F_{\rm{Contact}} = \frac{S_{D1, \rm{Contact}} \ \nu_{B} h}{2e^3 V_{0}\nu_{\rm QPC}(\nu_{B}-\nu_{\rm QPC})}.
\end{align}
Below we use the general formulas, Eqs.~(\ref{eq:noiseD1}) and (\ref{eq:FanofactorQPC}), in order to evaluate the noise and the corresponding Fano factor 
 for a specific class of filling factors $\nu_{\rm B}$ and $\nu_{\rm QPC}$.

\begin{figure}[t!]
\captionsetup[subfigure]{position=top,justification=raggedright}
\subfloat[]{
\includegraphics[width =0.85\columnwidth]{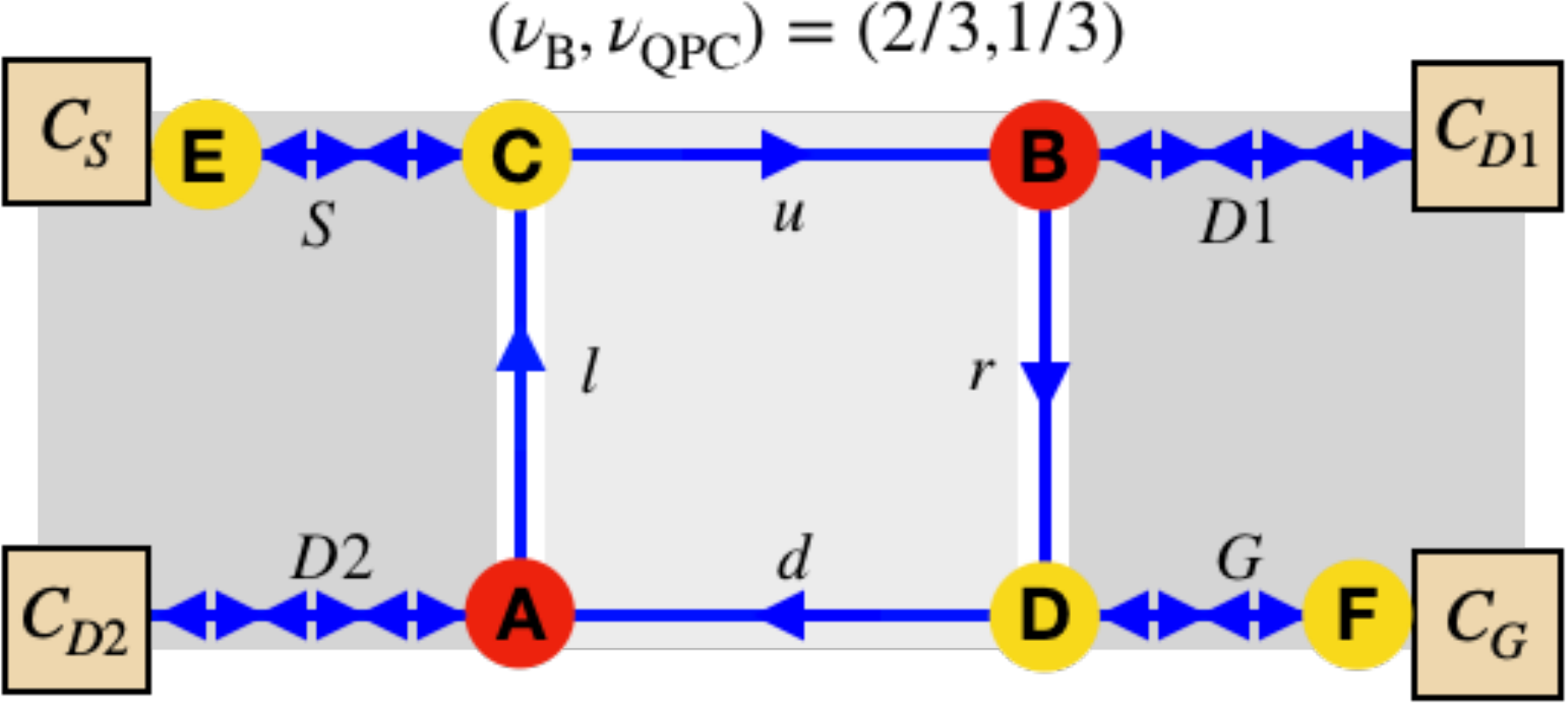}
\label{fig:Heat23}}
\\[-0.0cm]
\subfloat[]{
\includegraphics[width=0.85\columnwidth]{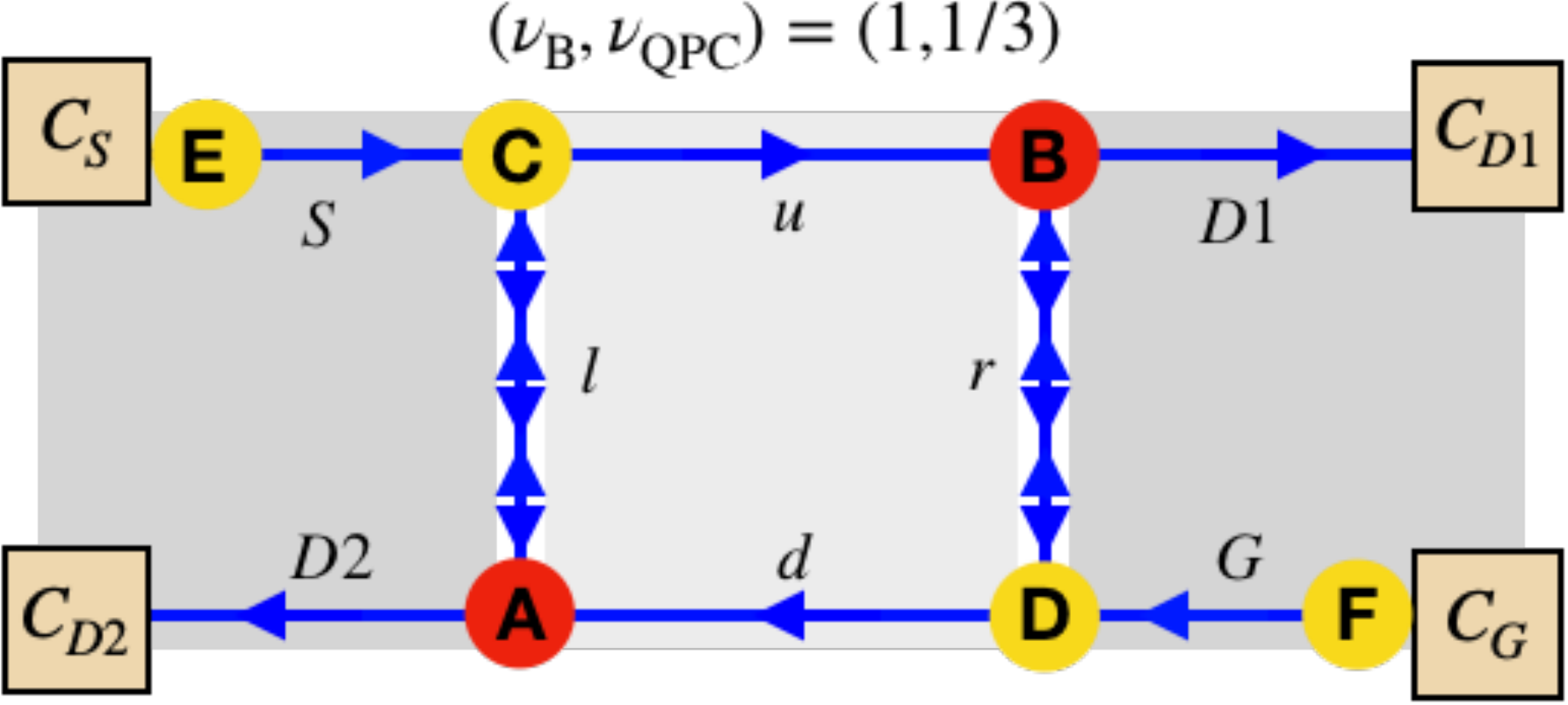}
\label{fig:Heat1}}
\\[-0.0cm]
\subfloat[]{
\includegraphics[width =0.85\columnwidth]{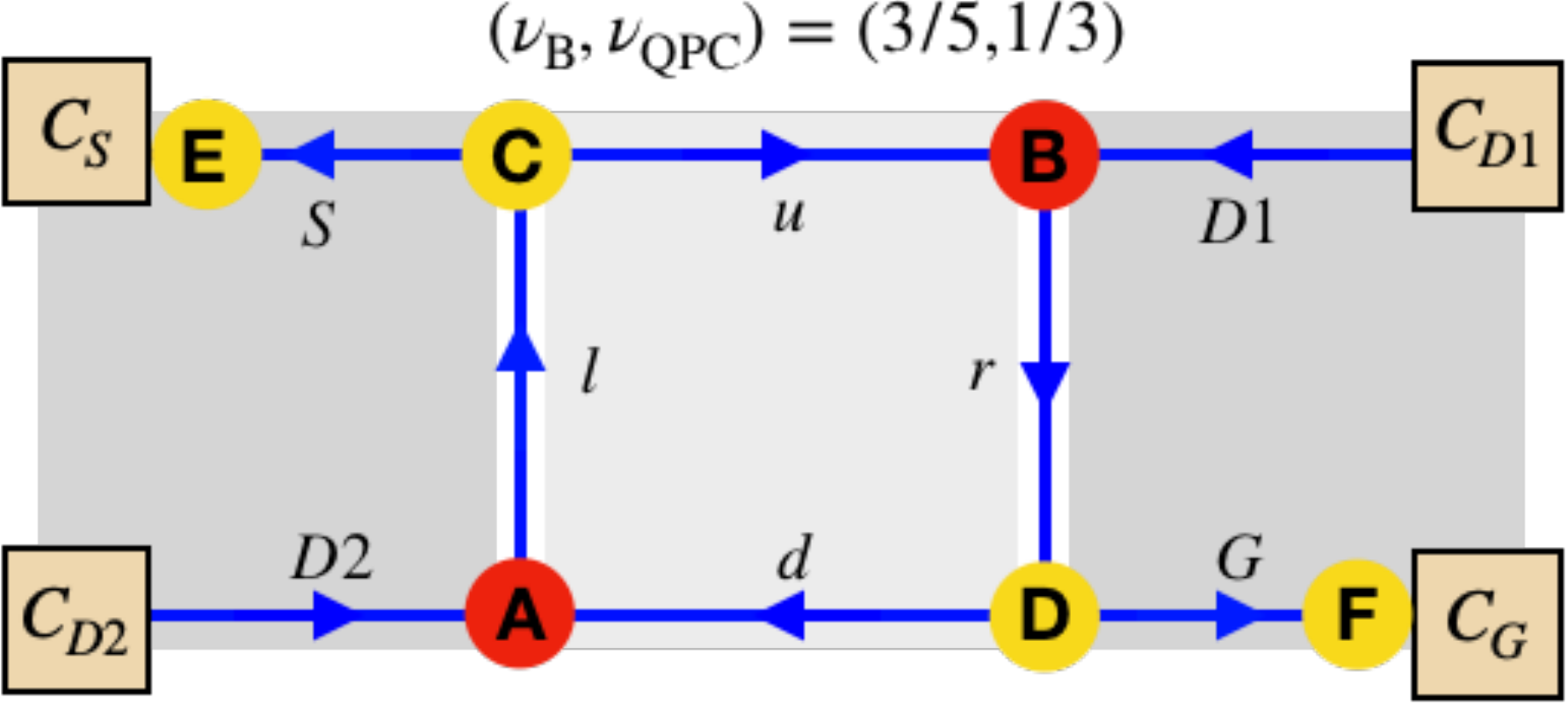}
\label{fig:Heat35}}
\caption{Edge heat propagation (blue arrows) in the incoherent QPC geometry for various choices of $\nu_{B}$ and $\nu_{\rm QPC}=1/3$. The contact $C_{S}$ is voltage biased while $C_{G}$ is grounded.  The chiral nature of the edge ensures that heat is generated only due to voltage drops at the hot spots (red dots, $A$ and $B$). In the steady state, this heat generates a rise in the local temperature at the noise spots (yellow dots, $C-F$). This local temperature increase, which depends on the nature of heat propagation, in turn determines the characteristics of the generated shot noise. (a) $\nu_{B}=2/3$. Heat propagation is diffusive along the outer arms  (double headed arrows), antiballistic in the line junctions, and ballistic on the upper and lower edge. Heat exits the system from all four contacts. (b) $\nu_{B}=1$.  Heat propagation is diffusive in the line junctions but ballistic everywhere else. If leakage of heat to the bulk is negligible, heat only exits the system from contacts $C_{D1}$ and $C_{D2}$.  (c) $\nu_{B}=3/5$. Heat propagation is antiballistic along the outer arms and the line junctions but ballistic on the upper and lower edges at the low density region. Heat can only leave the system via contacts $C_{S}$ and $C_{G}$. }
\label{fig:HeatPropagation}
\end{figure}

\subsection{Noise for the (AB, AB, B) heat transport configuration}
\label{sec:noise_subsectionC}
As an application of Eq.~\eqref{eq:FanofactorQPC}, we now focus on the QPC configuration where the heat propagation is antiballistic along the outer arms and the line junctions, but is ballistic along the upper and lower edges in the low density region (see Fig.~\ref{fig:Heat35}). Specific examples that we consider are $(\nu_{B}, \nu_{\rm{QPC}}) = (3/5, 1/3)$ and $(4/7, 1/3)$ which were both studied experimentally in Ref.~\onlinecite{Bhattacharyya2019}. As specified in Sec.~\ref{sec:QuantNoise}, the noise is constant for such combinations, i.e. it does not depend on any length scales in the system. 

Energy conservation at the hot spots leads to 
\begin{subequations}
\begin{align} \label{eq:energyconservation}
&P_{A} + \frac{\pi^2 (k_B T_{D})^2}{6h} |\nu_{Q, \rm{QPC}}|  \nonumber \\ 
&= \frac{\pi^2 (k_B T_{C})^2 }{6h} ( |\nu_{Q, \rm{QPC}}| + |\nu_{Q, B}| ) , \\
&P_{B} + \frac{\pi^2 (k_B T_{C})^2}{6h} |\nu_{Q, \rm{QPC}}|  \nonumber \\
&= \frac{\pi^2 (k_B T_{D})^2}{6h} ( |\nu_{Q, \rm{QPC}}|  + |\nu_{Q, B}| ),
\end{align}
\end{subequations}
where we have used that the dissipated power is transported away ballistically or antiballistically. The dissipated powers, $P_A=P_B$ due to the voltage drops are given by Eqs.~\eqref{eq:PQPC}. The temperatures at the noise spots $C$ and $D$ are then obtained as
\begin{align}
T_{C} = T_{D} = \frac{eV_0}{\pi k_B} \sqrt{\frac{3\nu_{\rm{QPC}}(\nu_{B}-\nu_{\rm{QPC}})}{\nu_{B} |\nu_{Q, B}|}}.
\end{align}
Then, $F_{\rm{QPC}}$, defined in Eq.~\eqref{eq:FanofactorQPC}, reads
\begin{align}
\label{eq:FQPC}
F_{\rm{QPC}}= \frac{2}{\pi} \sqrt{\frac{3\nu_{\rm{QPC}}(\nu_{B}-\nu_{\rm{QPC}})}{\nu_{B}|\nu_{Q,B}|}}.
\end{align}
We emphasize that the contribution $F_{\rm{QPC}}$ is topological in the sense that it is expressed solely through topological invariants $\nu$ and $\nu_Q$ (cf. Eqs.~\eqref{eq:G} and~\eqref{eq:GQ}) of the device edge segments.

Next, we compute the part $F_{\rm{Contact}}$ of the Fano factor. The noise generated in the outer arms is found according to Eq.~\eqref{eq:Noise2Modes} as
\begin{align}
\label{eq:NoiseExp2}
\overline{(\delta I_S)^2} &= \overline{(\delta I_G)^2} \nonumber  \\
& \simeq \frac{2e^2}{h l_{\rm{eq}}}\frac{ \nu_-}{ \nu_+} \nu_{B} \int_0^{L_{\rm Arm}} dx\;\frac{e^{-\frac{2x}{l_{\rm{eq}}}}k_B\left(T_+(x) +T_-(x) \right)}{(1-e^{-\frac{L}{l_{\rm{eq}}}}\nu_-/\nu_+)^2}.
\end{align}
Here $\nu_{+ (-)}$ is the total filling factor of the downstream (upstream) mode in the outer arms and thus $\nu_{B} = \nu_{+} - \nu_{-}$.
To obtain the temperature profiles, we solve Eq.~\eqref{eq:TemperatureEquation} with boundary conditions $T_+(0)=0$, and $T_-(L_{\rm Arm})=T_C$. As there are no voltage drops along the edge segments $C_{S}-C$ and $C_{G}-D$, the Joule heating contribution (see Eq.~\eqref{eq:Vdiff2}) vanishes. In the limit $l_{\rm{eq} }\ll L_{\rm Arm}$, we find
\begin{subequations}
	\begin{align}
		& k_B^2T^2_+(x) = k_B^2T_C^2\frac{n_u-n_ue^{-\alpha x/l_{\rm eq}}}{n_u-n_d e^{-\alpha L/l_{\rm eq}}}, \\
		& k_B^2T^2_-(x) = k_B^2T_C^2\frac{n_u-n_de^{-\alpha x/l_{\rm eq}}}{n_u-n_d e^{-\alpha L/l_{\rm eq}}}. 
	\end{align}
\end{subequations}
where we have introduced the parameter $\alpha \equiv -(n_d-n_u)\gamma \nu_+ \nu_-/\nu_{\rm B}$. Using these general profiles in Eq.~\eqref{eq:NoiseExp2}, we find the asymptotics
	\begin{align}
	\label{eq:EFContr}
	& \hspace*{-0.5cm} \overline{(\delta I_S)^2}=\overline{(\delta I_G)^2} \simeq \frac{e^2}{h} \frac{\nu_B \nu_+k_B^2 T_C^2}{\nu_-}
	 \notag \\ 	 \times
	&  \left[\frac{\sqrt{\pi } \Gamma \left(\frac{2+\alpha}{\alpha }\right)}{2 \Gamma \left(\frac{3}{2}+\frac{2}{\alpha }\right)}+\,_2F_1\left(-\frac{1}{2},\frac{2}{\alpha };\frac{2+\alpha}{\alpha};\frac{n_d}{n_u}\right)\right],
	\end{align}
where $\Gamma(c)$ is the gamma function and $_2F_1(a,b;c;z)$ is the hypergeometric function. 

\begin{figure}[t!]
\includegraphics[width=0.99\columnwidth]{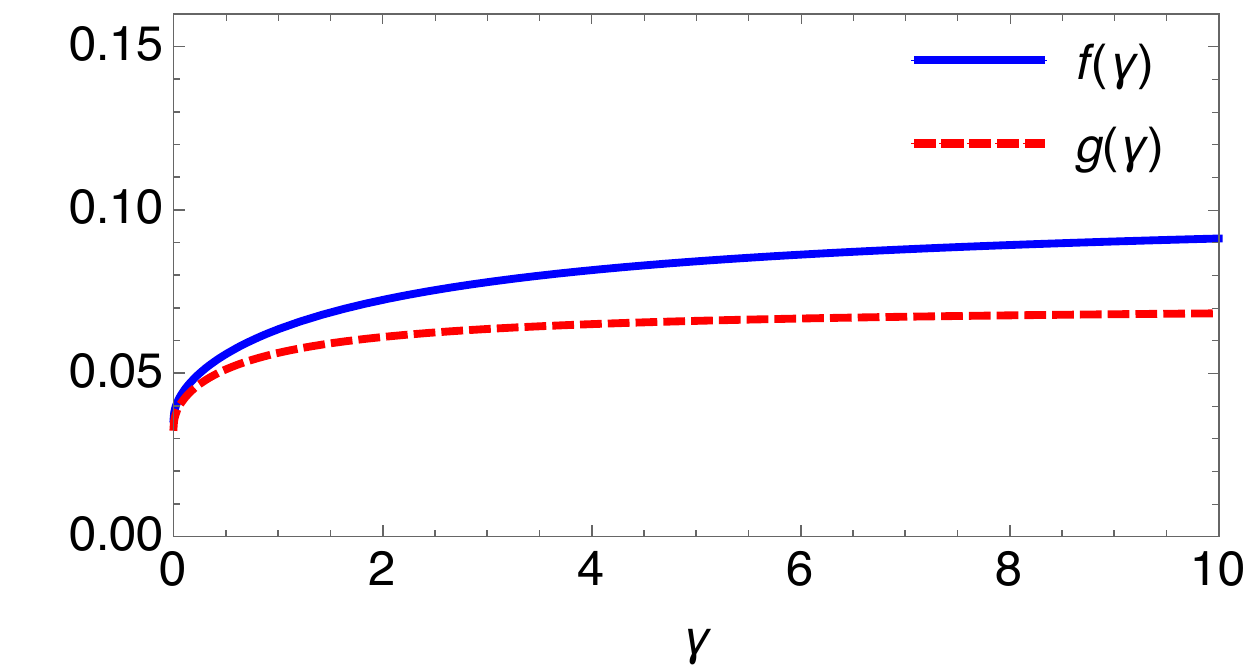}\caption{\label{fig:NoiseFunctions} The dimensionless functions $f(\gamma)$ and $g(\gamma)$ which determine the dependence of shot noise for the transport through a QPC with $(\nu_{B}, \nu_{\rm{QPC}}) = (3/5, 1/3)$ and $(4/7, 1/3)$ on the parameter $\gamma$ controlling the violation of the Wiedemann-Franz law in the inter-mode tunneling. The exact functional expressions are given in Eqs.~\eqref{eq:fgamma} and~\eqref{eq:ggamma}.}
\end{figure}
We next apply this formula to the combination $(\nu_{B}, \nu_{\rm{QPC}}) = (3/5, 1/3)$. The microscopic composition of $\nu_{\rm B}$ is $\nu_+ = 1$, $\nu_- = 2/5$, $n_u = 1$, $n_d=2$, and $\nu_{Q,\rm{B}}=-1$. The contribution  $F_{\rm{QPC}}$ to the Fano factor, Eq.~\eqref{eq:FQPC}, then becomes 
$$F_{\rm{QPC}}\approx 0.42,$$ 
while Eq.~\eqref{eq:EFContr} reduces to
\begin{align}
\label{eq:fgamma}
	&\overline{(\delta I_S)^2} = \overline{(\delta I_G)^2} \simeq \frac{e^3V_0}{h}
	\notag \\
	& \times \frac{4 \left(\frac{\sqrt{\pi } \Gamma \left(\frac{\gamma +3}{\gamma }\right)}{2 \Gamma \left(\frac{3}{2}+\frac{3}{\gamma }\right)}+\, _2F_1\left(-\frac{1}{2},\frac{3}{\gamma };\frac{\gamma +3}{\gamma };\frac{1}{2}\right)\right)}{25 \pi  } \equiv \frac{e^3 V_0}{h} f(\gamma).
\end{align}
The only remaining parameter is the Wiedemann-Franz parameter $\gamma$ (see Sec.~\ref{sec:Model}) which depends on the interaction between the modes and is unknown. We plot $f(\gamma)$ in Fig.~\ref{fig:NoiseFunctions}. For $\gamma=1$, we obtain 
$$\overline{(\delta I_S)^2} = \overline{(\delta I_G)^2}=f(1)\approx 0.063,$$ 
which leads to 
$$F_{\rm{Contact}} \approx 0.032.$$ 
The total Fano factor is then
\begin{equation}
	F_{3/5,1/3} \approx 0.45,
	\label{eq:F3513}
\end{equation}
which is constant as expected.

For the configuration $(\nu_{B}, \nu_{\rm{QPC}}) = (4/7, 1/3)$ the analysis is carried out in the same way. We have in this case $\nu_+ = 1$, $\nu_- = 3/7$, $n_u = 1$, $n_d=3$, and $\nu_{Q,\rm{B}}=-2$. Equation \eqref{eq:FQPC} then yields
$$F_{\rm{QPC}}\approx 0.29,$$
while Eq.~\eqref{eq:EFContr} leads to
\begin{align}
\label{eq:ggamma}
	&\overline{(\delta I_S)^2} = \overline{(\delta I_G)^2} \simeq \frac{e^3V_0}{h}	
	\notag \\ 
	& \times \frac{2\sqrt{10}\left(\frac{\sqrt{3\pi } \Gamma \left(1+\frac{4}{3\gamma}\right)}{4 \Gamma \left(\frac{3}{2}+\frac{4}{3\gamma }\right)}+\frac{\sqrt{3}}{2}\, _2F_1\left(-\frac{1}{2},\frac{4}{3\gamma };1+\frac{4}{3\gamma};\frac{1}{3}\right)\right)}{49 \pi } \notag \\ & \equiv \frac{e^3 V_0}{h} g(\gamma).
\end{align}
The function $g(\gamma)$ is plotted in Fig.~\ref{fig:NoiseFunctions}. With $\gamma =1$, we obtain 
$$F_{\rm{Contact}} \approx 0.027. $$
Hence, 
\begin{equation}
	F_{4/7,1/3} \approx 0.32.
	\label{eq:F4713}
\end{equation}
The above values of the contact contribution $F_{\rm{Contact}}$ to the Fano factor  are obtained for $\gamma=1$. Varying $\gamma$ does not modify the order of magnitude of $F_{\rm{Contact}}$ but can change it within a factor $\sim 2$ (see Fig.~\ref{fig:NoiseFunctions}). 

Experimentally found values of the noise for the $(\nu_{B}, \nu_{\rm{QPC}}) = (3/5, 1/3)$ and $(4/7, 1/3)$ plateaus are $F\approx 0.6$ and $F\approx 0.56$, respectively~\cite{Bhattacharyya2019}.  These values are comparable with but somewhat larger than our results~\eqref{eq:F3513} and~\eqref{eq:F4713}. 

We end this section by discussing the effect of possible leakage of the heat to the bulk of the system.  Let us assume that the length scale over which all the heat is transferred into the bulk is much smaller than $L_{\rm{Arm}}$ but much larger than $L_{\rm{QPC}}$. In this case, the fluctuations $\overline{(\delta I_{S})^2} $ and $\overline{(\delta I_{G})^2}$ vanish since no heat generated at the hot spots reaches the noise spots $E$ and $F$ but instead leaks into the bulk. Then, the Fano factor $F$ is given entirely by the topological quantity $F_{\rm QPC}$.  We also note that, even if the heat leakage is fully absent, the contact contribution $F_{\rm{Contact}}$ turns out to be numerically much smaller than $F_{\rm QPC}$.  Therefore, $F$ remains numerically close to $F_{\rm QPC}$ independently of the degree of heat leakage to the bulk within the distance $L_{\rm{Arm}}$.

\section{\label{sec:Discussion}Discussion}
\subsubsection{Noisy QPC conductance plateaus: comparison with experiments}
\label{sec:experiment}
Our model, which assumes  the incoherent transport regime, i.e., full equilibration along all edge segments, predicts formation of quantized plateaus in the transport through a QPC at fractions $\nu_{\rm QPC}<\nu_{\rm B}$ corresponding to stable FQHE states. We emphasize that this plateau formation phenomenon is very general and robust. In particular, the plateau formation does not require any connection between the structure of the edges at fractions $\nu_{\rm QPC}$ and $\nu_{\rm B}$. Neither does it depend on any proximity of the system to some RG fixed point. The only assumption is the incoherent regime. The prediction of this model on the ubiquity of such quantized conductance plateaus is in agreement with experimental findings of Refs.~\onlinecite{Ando1998,Bid2009,Sabo2017,Bhattacharyya2019} which reveal an abundance of plateaus for various combinations of $\nu_{\rm QPC}$ and $\nu_{\rm B}$. In particular, the fact that the plateaus occur also for $\nu_{\rm B} =1$ demonstrates  that no special properties of the $\nu_{\rm B}$ edge is required for this phenomenon. We are not aware of any other mechanism that would explain formation of plateaus for a generic pair of fillings $\nu_{\rm QPC}$ and $\nu_{\rm B}$. 

The experiment~\cite{Bhattacharyya2019} reported that the most prominent $G=(1/3) e^2/h$ plateau is observed only for $\nu_{\rm B}<5/3$ and disappears for higher $\nu_{\rm B}$. 
How can this be reconciled with our theory?  We can explain this observation by assuming that the equilibration length increases strongly for larger $\nu_{\rm B}$. A recent measurement~\cite{Lin2019} supports this assumption, where the equilibration length between edge channels from different Landau levels was estimated to be an order of magnitude longer than that between channels within a single Landau level.
  We can therefore argue that the absence of the $G=(1/3) e^2/h$ plateau for larger filling factors is due to insufficient equilibration: the assumption of the incoherent regime, $l_{\rm eq}\ll L_{\rm{QPC}}$, gets violated. In addition, it is plausible that the stability of the $\nu_{\rm QPC} = 1/3$ state in the QPC region is reduced with weaker magnetic fields (i.e. larger $\nu_{\rm B}$).  As we discuss below, violation of the condition of incoherent transport regime at larger $\nu_{\rm B}$ is also supported by the observation of Mach-Zehnder interference at these filling fractions. 
    
The second prediction of our model is that the quantized plateaus in the QPC transport are in general noisy.  This is in qualitative agreement with the findings of Ref.~\onlinecite{Bhattacharyya2019}. At the same time, the values of the Fano factor $F$ that we find differ from the relation $F = \nu_{\rm B}$ proposed in Ref.~\onlinecite{Bhattacharyya2019}. In fact, for several of combinations ($\nu_{\rm B}$, $\nu_{\rm QPC}$) studied experimentally, we find constant Fano factors $F$ with numerical values not far from those reported in Ref.~\onlinecite{Bhattacharyya2019} (see Sec.~\ref{sec:noise_subsectionC}). On the other hand, for other combinations of filling factors, we find Fano factors that depend on ratios of length scales. Further work, both experimental and theoretical, should help to better understand the difference in the values of $F$ between the experiment of  Ref.~\onlinecite{Bhattacharyya2019} and our theory. Some promising directions are discussed in Sec.~\ref{sec:Summary}. 

We have also studied the transport through structure with two identical QPCs (see Sec.~\ref{sec:DQPCresults}).  Our theory predicts conductance plateaus also in this case, with the same value of the conductance as in a single QPC for two closely located QPCs and with a distinct value for a sufficiently large distance between QPCs. These results are in full agreement with the experimental findings of Ref.~\onlinecite{Sabo2017}. 
  
\subsubsection{Mach-Zehnder interferometry}
We further discuss the Mach-Zehnder interferometry measurements performed in Ref.~\onlinecite{Bhattacharyya2019}. The fundamental requirement for interference is phase coherence. An important observation of Ref.~\onlinecite{Bhattacharyya2019} is that there are clear ``anti-correlations'' between the observation of Mach-Zehnder interference and the the quantized plateaus in the QPC transport:  whenever the (most prominent) $1/3$ plateau is formed, the Mach-Zehnder interference is strongly suppressed. This observation is fully consistent with our theory of QPC plateaus as resulting from the incoherent transport. Indeed, the loss of coherence in this transport regime should lead to vanishing of Mach-Zehnder interference in agreement with experiment.  Conversely, the disappearance of $1/3$ plateau at higher bulk filling factors, $\nu_{\rm B} > 5/3$, indicates that the transport is (at least partly) coherent, in consistency with observation of Mach-Zehnder interference at these values of $\nu_{\rm B}$. 

\subsubsection{Relation to edge reconstruction}
\begin{figure}[t!]
\includegraphics[width=0.99\columnwidth]{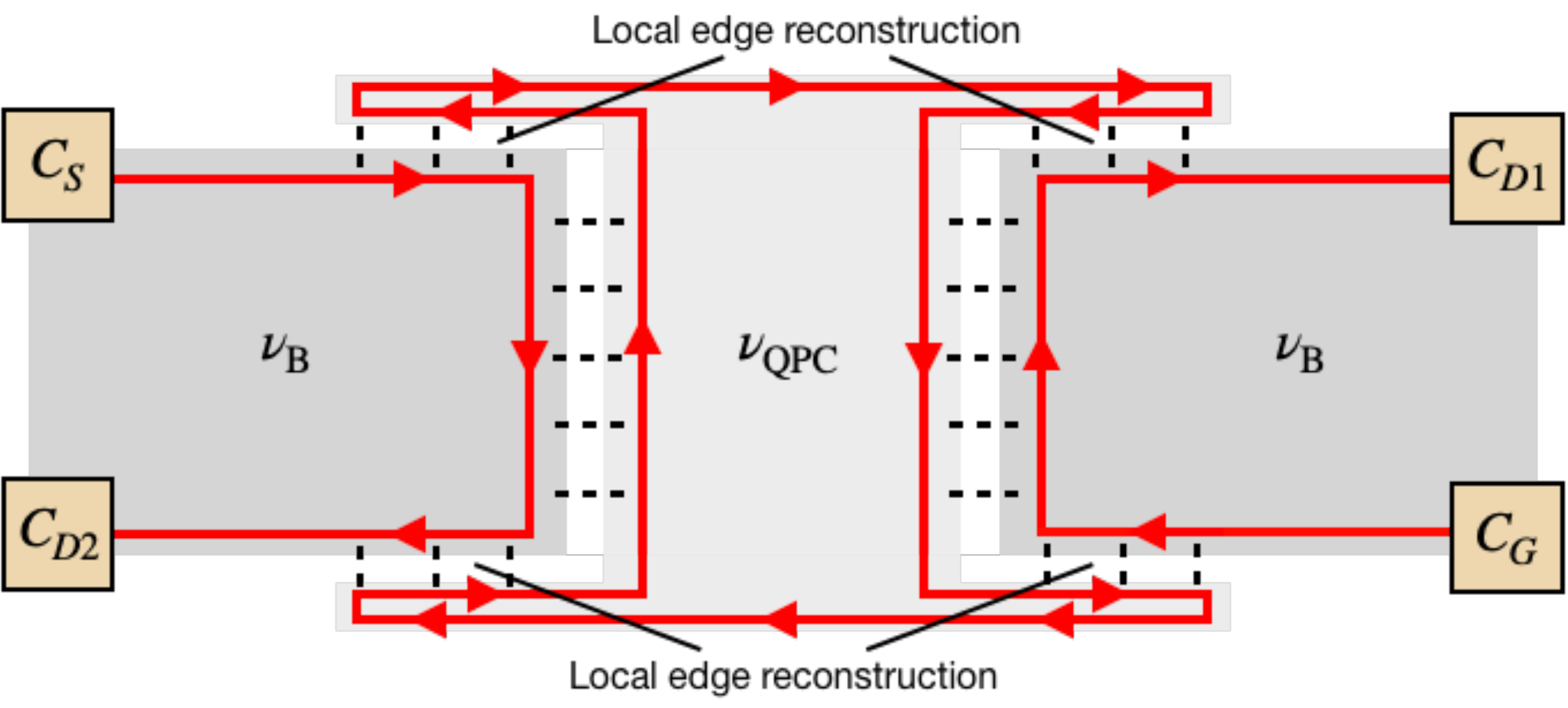}\caption{\label{fig:Edge Reconstruction} 
Deformation of the shape of the low-density area in Fig.~\ref{fig:QPCSetup} yields a configuration that can be viewed as a ``local edge reconstruction'' of the $\nu_B$ edge by  a pair of counterpropagating $\nu_{\rm QPC}$ modes.}
\end{figure}

It may be instructive to draw a certain parallel between our model and the picture of edge reconstruction due to a soft confinement potential~\cite{Meir1994,Wang2013,Sabo2017}. Specifically, let us somewhat deform the low-density area in Fig.~\ref{fig:QPCSetup}, bringing it to the form shown in Fig.~\ref{fig:Edge Reconstruction}.  We see that this configuration can be viewed as a ``local edge reconstruction'' of the $\nu_B$ edge by a pair of counterpropagating modes with filling factor discontinuities $\delta \nu_{1,2}=\pm\nu_{\rm QPC}$. 
We emphasize, however, crucial differences between the mechanism of formation of the QPC conductance plateau in our work and that proposed in Refs.~\onlinecite{Wang2013,Sabo2017}. Contrary to these works,  we do not assume any reconstruction of the $\nu_{\rm B}$ edge as such (our analysis is instead insensitive to the presence or absence of such reconstruction). The additional  $\nu_{\rm QPC}$ edge appears only in the vicinity of the QPC due to depletion of the density. Moreover, we do not assume any coherent renormalization towards specific fixed points; our analysis is fully general in this respect as well. As has been already emphasized in the beginning of the paper, the only assumption is the incoherent transport regime, i.e., full equilibration. 
  \subsubsection{Partially coherent regime}  
  Finally, we briefly discuss the QPC transport in the partially coherent regime, $L_{\rm QPC}\ll l_{\rm eq} \ll L_{\rm Arm}$ (see Fig.~\ref{fig:LengthScales}). In this case, there is no generic reason for the formation of conductance plateaus with $G = \nu_{\rm QPC} (e^2/h).$ To illustrate this point, we consider the specific case $\nu_{\rm B} = 1$ and $\nu_{\rm QPC} = 1/3$, which is one of the prominent examples in which a well developed plateau was observed experimentally~\cite{Bhattacharyya2019}. The corresponding results for the conductance $G$ can be inferred from the analysis in Sec.~6.2 of Ref.~\onlinecite{Protopopov2017}. That paper studied a slightly different experimental setup (with an intermediate $2/3$ region), which, however, was described by essentially the same effective model:  a floating $1/3$ mode forming two line junctions with $1$-modes. It is easy to see that our conductance $G$ can be obtained by subtracting from unity the conductance $G$ of Ref.~\onlinecite{Protopopov2017}. We thus find the following results. If the random tunneling along the line junctions is weak, the conductance $G$ is zero, up to a small correction.  If the interface at the line junctions is renormalized towards the KFP fixed point (which happens for strong random tunneling), the conductance shows strong mesoscopic fluctuations in the range $0 < G / (e^2/h) < 1$, i.e. the value of the conductance depends on the microscopic details such as the specific realization of disorder or the gate potential.
  It would be very interesting to observe such a regime experimentally (see also a discussion in Sec.~\ref{sec:Summary}) but presumably this requires very low temperatures.
  In none of the described situations a $G = (1/3)e^2/h$ plateau can emerge.  Thus, the formation of this plateau unambiguously points towards the incoherent transport regime. 
  
  For other pairs of filling factors $\nu_{\rm B}$ and $\nu_{\rm QPC}$, the analysis of the partially incoherent regime is more involved. Furthermore, one can imagine situations with hierarchy of equilibration lengths, which will produce a wealth of possible types of behavior of $G$. However, a paradigmatic example of $\nu_{\rm B} = 1$ and $\nu_{\rm QPC} = 1/3$ shows that, if the mechanism for formation of plateaus is the same for various  pairs $\nu_{\rm B}$, $\nu_{\rm QPC}$ (which is natural to expect), this should be the incoherent (fully equilibrated) transport regime.

\section{\label{sec:Summary}Summary and outlook}
In this paper we studied incoherent transport and shot noise in QPC setups in the FQH regime. We modelled the QPC as 
a low density region with $\nu_{\rm QPC}<\nu_{\rm B}$  inside a FQH state with filling factor $\nu_{\rm B}$, depicted in Fig.~\ref{fig:QPCSetup}. Effectively, the QPC is then equivalent to two line junctions in series. The generic incoherent regime was implemented by assuming full equilibration for both heat and charge: the characteristic equilibration length $l_{\rm eq}$ was assumed to be much smaller than the physical lengths of the edge segments.

Our main results are as follows. First, we explained  the formation of QPC conductance plateaus for a wide variety of FQH states. As a prominent example, our theory explains 
the fractional $G=(1/3) e^2/h$ conductance plateaus for the integer state $\nu_{\rm B}=1$. Such a plateau may look rather unexpected at first sight since the integer $\nu_{\rm B}=1$ edge does not contain any fractional modes. We have further studied conductance plateaus in the double-QPC setup assuming two equivalent QPCs. When two QPC are closely located, they serve as a single QPC. For a sufficiently large distance between QPC, our theory predicts plateaus with distinct values of the conductance, again in agreement with experiment. 

Secondly, we explored the generation of shot noise in the single-QPC geometry. We showed that, depending on the values of $\nu_{\rm B}$ and $\nu_{\rm QPC}$, the system falls into one of  13 ``universality classes'' with different combinations of heat transport on segments. We have complemented this topological classification of shot noise in QPC transport by the analysis of the noise in various classes. Apart from one class, where the noise is exponentially suppressed in the incoherent regime, the QPC system is characterized by a rather strong noise. This important conclusion of the theory that the conductance plateaus are in general noisy is in qualitative agreement with experiment. The specific behavior of the effective Fano factor $F$ depends on the ``topological universality class'' of the system. In a number of important classes, the noise is constant (i.e, $F$ is a constant of order unity). Up to a numerically small contribution originating from contacts, this constant is a topological invariant, i.e., it is fully determined by the topology of the $\nu_{\rm B}$ and $\nu_{\rm QPC}$ states. At the same time, the constant $F$ is irrational and cannot be interpreted as an effective charge of quasiparticles. In other classes, the noise shows a weak (square-root) power-law dependence on relevant lengths characterizing the system. This includes also situation in which the noise can become super-poissonian ($F>1$) if the heat leakage to the bulk (neglected in our analysis) is sufficiently weak. 

Finally, we elaborated on the visibility in the recently performed Mach-Zehnder interferometry. We argued that the loss of coherence due to strong equilibration explain the drastic loss of visibility for lower filling factors $\nu_{\rm B}$. This explains in particular why the strong suppression of the Mach-Zehnder interference is experimentally correlated with the appearance of the $1/3$ plateau in the QPC transport. 

We hope that this paper will stimulate further experimental and  theoretical investigations of transport through various setups in the FQH regime. Let us discuss a few prospective research directions. 

On the theory side, one particularly interesting extension of our work is to study the proposed QPC model with non-Abelian filling factors, most notably $\nu_{\rm B}=5/2$. The exact edge structure of this state is not known, and we envision that QPC experiments could provide important input for determining it. 
 Another direction is a microscopic study of the partially coherent regime for various values of $\nu_{\rm B}$ and $\nu_{\rm QPC}$.  Further, it is important to study the Mach-Zehnder interferometry in various regimes, from fully coherent to fully incoherent. This is expected to be very useful towards the goal of constructing a Mach-Zehnder interferometer device operating in the FQH regime.  
 
 On the experimental side, an important challenge is to observe the crossover from the coherent to incoherent regime in the QPC transport. A recent experimental breakthrough 
 based on a specially designed double-well structure has allowed observation of such a crossover in the two-terminal conductance of a $2/3$ edge~\cite{Cohen2019}. We hope that this technological and experimental progress will permit also to explore the coherent-to-incoherent crossover in transport through a QPC. While various combinations of $\nu_{\rm B}$ and $\nu_{\rm QPC}$ are of great interest, such an experiment for the paradigmatic case $\nu_{\rm B} = 1$ and $\nu_{\rm QPC} =1/3$ would be a very important reference point. Furthermore, a systematic experimental study of the dependence of the noise in the QPC transport with various pairs ($\nu_{\rm B}$, $\nu_{\rm QPC}$) with system parameters (such as the temperature, which affects the equilibration length) would be very desirable. It would be extremely useful if such an investigation is conducted parallel to the study of the noise in two-terminal conductance of an edge \cite{Park2019,Spanslatt2019}, which is a simpler setup. The classification of the noise in the two-terminal transport \cite{Spanslatt2019} and in the QPC transport (the present work) can be used to probe the character of the heat transport in these devices. 
 Another attractive experimental direction is to use local thermometry~\cite{Halbertal2016} to probe the heat transport, and in particular to image the two hot spots predicted by our model (see Fig.~\ref{fig:DQPCa}). 

\begin{acknowledgments}
We thank R. Bhattacharyya, M. Heiblum,  C. Hong, A. Rosenblatt, and B. Rosenow for useful discussions. C.S.,  Y.G., and  A.D.M. acknowledge support by DFG Grant No. MI 658/10-1.  
\end{acknowledgments}

\appendix

\section{Hot spots in the vicinity of drains}
\label{sec:AppHotspots}
In this Appendix, we outline the experimental implementation of the drain contacts and show that due to their specific configuration, heat generated at the drains cannot flow back to the QPC region and contribute to the noise generation. 

\begin{figure}[t]
\includegraphics[width=0.8\columnwidth]{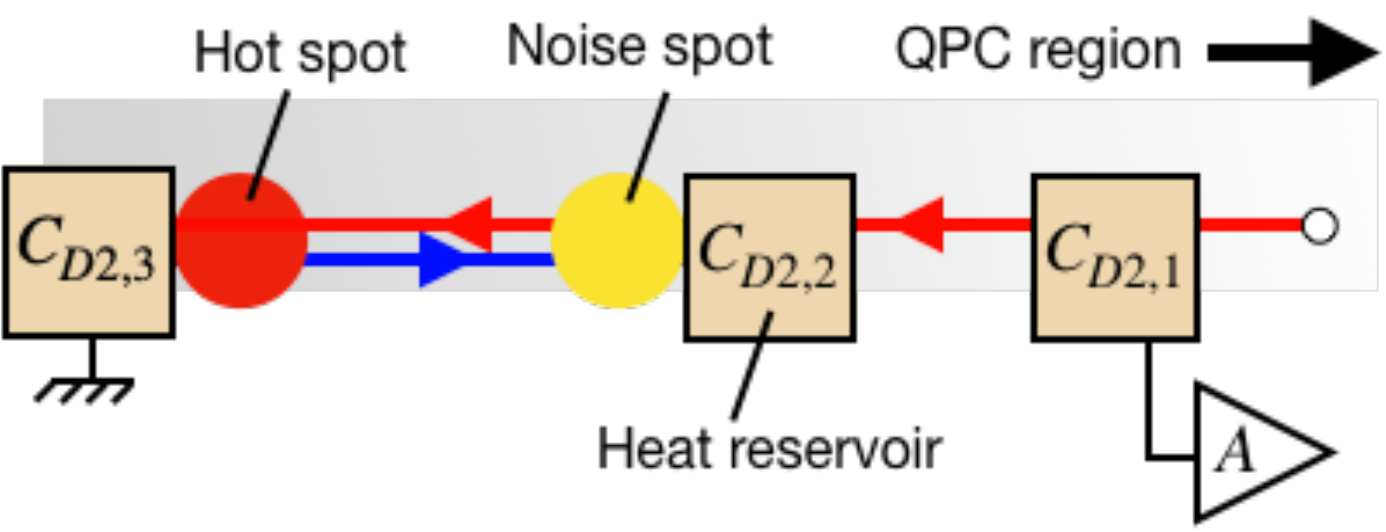}\caption{\label{fig:drainContacts} Schematic diagram for the $C_{D2}$ drain contact specifying the experimental configuration in Refs.~\onlinecite{Bid2010, Bhattacharyya2019}. This setup prevents that any heat generated at the drain flows back to the QPC region, where it would contribute to the noise. 
The drain (simply depicted by a single contact in previous figures) consists of three individual contacts: a floating contact $C_{D2, 1}$ measuring the noise from voltage fluctuations, another floating contact $C_{D2, 2}$ acting as a heat reservoir, and 
a contact $C_{D2, 3}$ connected to the ground. The net electrical current impinges from the right hand side of $C_{D2, 1}$ through a QPC (not shown here). The charge (heat) propagation direction is drawn with red arrows (a blue arrow). The hot spot and a possible noise spot are denoted as red and yellow dots, respectively. A similar configuration holds for the other drain $C_{D1}$.}
\end{figure}

We consider the drain contacts fabricated in Refs.~\onlinecite{Bid2010, Bhattacharyya2019} (see Fig.~\ref{fig:drainContacts}). This type of drain (for simplicity depicted by a single contact in the previous figures) actually consists of three nearby contacts: a floating contact $C_{D2, 1}$ that serves to measure voltage noise (which can eventually be converted into current noise), another floating contact $C_{D2, 2}$ acting as a heat reservoir, and a third contact $C_{D2, 3}$ connected to ground. The electrical current arrives to the ground contact $C_{D2, 3}$ from the QPC region. The voltage drop and associated Joule heating, depicted as a red dot, occur only in the vicinity of the ground contact $C_{D2, 3}$. The generated heat can potentially propagate upstream (the blue arrow) to $C_{D2, 2}$ but cannot propagate beyond the heat reservoir $C_{D2, 2}$. Hence, no heat may flow back to the QPC and contribute to the generation of the noise discussed in the main text. We can therefore safely neglect the effect of heat generation at the drain contacts. 

We may, however, ask whether the generated heat can induce additional noise during propagation between the drain contacts. According to the noise-generating mechanism specified in Sec.~\ref{sec:Noise Mechanism}, thermally activated tunneling between edge channels (due to the heating along the edge) can excite particle-hole pairs. If only one of the constituents of such pair would reach the contact $C_{D2, 1}$ (at which the noise is measured), such a tunneling event would contribute to the noise. A possible noise spot is located in the the vicinity of the left hand side of $C_{D2, 2}$ (depicted as a yellow dot). If the length between $C_{D2,1}$ and $C_{D2, 2}$ is  larger than the equilibration length $l_{\rm{eq}}$, particles and holes generated in the process of tunneling at this would-be noise spot will actually all eventually flow downstream (to the ground contact $C_{D2, 3}$), so that no noise will be generated in $C_{D2, 1}$. The above condition, that the length between $C_{D2,1}$ and $C_{D2, 2}$ is  larger than $l_{\rm{eq}}$, was reasonably satisfied in these experiments. We therefore conclude that the specific contact configuration allows us to ignore the effect of the heating at the drain contacts.

\section{Hot spot temperatures, noise, and Fano factors for specific QPC configurations}
\label{sec:AppNoise}
This appendix supplements Sec.~\ref{sec:noise_subsectionC} where we studied noise 
for two configurations $(\nu_{B}, \nu_{\rm{QPC}}) = (3/5, 1/3)$ and $(4/7, 1/3)$, which both yield (AB, AB; B) heat transport configurations and constant noise. Here, we analyze the configurations $(\nu_{\rm B},\nu_{\rm QPC}) = (1,1/3)$ and $(2/3,1/3)$, which correspond to (B, D, B) and (D, AB, B) respectively, and give two distinct types of length-dependent noise in the incoherent limit (see Table \ref{tab:QPCQualitative}).

\subsection{$(\nu_{\rm B},\nu_{\rm QPC}) = (1,1/3)$}
\label{sec:s_113}
The bulk state $\nu_{\rm B}=1$ hosts only a single edge mode, and we  immediately find that $\overline{(\delta I_S)^2}=\overline{(\delta I_G)^2}=0$, since no partitioning can occur close to the contacts $C_S$ and $C_G$. According to Eq.~\eqref{eq:noiseD1}, what then remains to compute are the temperatures at the noise spots $C$ and $D$ (see Fig.~\ref{fig:Heat1}).

To this end, we first note that the heat transport is ballistic on the outer segments (i.e., those with lengths $L_{\rm{Arm}}$.) Since we take $\nu_{\rm{QPC}}=1/3$, the heat transport in the line junctions (with lengths $L_{\rm{QPC}}$) is diffusive. Assuming zero temperature in the contacts $C_S$ and $C_G$, we can write the heat currents $J_i$ along all edge segments $i$ as 
\begin{subequations}
\begin{align}
	&J_{S} = 0, \\
	&J_{u} = \frac{\kappa}{2} T^2_C, \\
	&J_{D1} = \frac{\kappa}{2} T^2_B, \\
	&J_{G} = 0, \\
	&J_{d} = \frac{\kappa}{2} T^2_D, \\
	&J_{D2} = \frac{\kappa}{2} T^2_A, \\
	&J_{l} = \frac{\tilde{l}\kappa(T_A^2-T_C^2)}{2L_{\rm{QPC}}} ,\\
	&J_{r} = \frac{\tilde{l}\kappa(T_B^2-T_D^2)}{2L_{\rm{QPC}}},
\end{align}
\end{subequations}
where $\kappa = \pi^2k_B^2/3h$ and $\tilde{l}$ is a characteristic diffusion length. From inspecting diffusive solutions of Eq.~\eqref{eq:TemperatureEquation}, we take $\tilde{l} = l_{\rm eq}(\nu_{\rm B}-\nu_{\rm QPC})/(\nu_{\rm B} \nu_{\rm QPC} \gamma)=2l_{\rm eq}/\gamma$. Conservation of energy implies
\begin{subequations}\label{eq:HeatEqSys1}
\begin{align}
	&J_{D2}+J_{l} = P_{A}+J_{d},\\ 
	&J_{G}+J_{d} = J_{r}, \\ 
	&J_{D1}+J_{l} = P_{B}+J_{u}, \\ 
	&J_{S}+J_{u} = J_{l},
\end{align} 
\end{subequations}
where $P_{A}$ and $P_{B}$ are the dissipated powers at the two hot spots $A$ and $B$. These powers are equal and determined by Eq.~\eqref{eq:DissPower}. Inserting $\nu_{\rm{B}}=1$ and $\nu_{\rm{QPC}}=1/3$ into that equation yields
\begin{equation}
	P_{A} = 	P_{B} = \frac{e^2}{h}\frac{V_{0}^2}{9}.
\end{equation}
By solving the system~\eqref{eq:HeatEqSys1} for $T_C$ and $T_D$, we find
\begin{equation}
\label{eq:TC1}
	k_B^2T^2_C = k_B^2T^2_D = \frac{4e^2 l_{\rm eq}V_0^2}{3\pi^2(2l_{\rm eq}+\gamma L_{\rm{QPC}})}\approx \frac{4e^2 l_{\rm eq}V_0^2}{3\pi^2\gamma L_{\rm{QPC}}},
\end{equation}
where we used $l_{\rm eq}\ll L_{\rm{QPC}}$.

Combining Eqs.~\eqref{eq:TC1}, \eqref{eq:noiseD1} and~\eqref{eq:FanofactorQPC}, we finally obtain the Fano factor for $(\nu_{\rm B},\nu_{\rm QPC}) = (1,1/3)$:
\begin{equation}
\label{eq:FT}
	F_{1,1/3} = \frac{k_B(T_C+T_D)}{eV_0} \approx 0.74 \sqrt{\frac{l_{\rm eq}}{\gamma L_{\rm{QPC}}}}.
\end{equation}
This result is in agreement with qualitative discussion in Sec.~\ref{sec:QuantNoise}, see Eq.~\eqref{eq:S_BBD}.   The Fano factor shows the  
$(l_{\rm eq} / L_{\rm{QPC}})^{1/2}$ length dependence  due to the diffusive nature of the heat propagation in line junctions. 

\subsection{$(\nu_{\rm B},\nu_{\rm QPC}) = (2/3,1/3)$}
\label{sec:s_2313}
Since the state $\nu_{\rm{B}}=2/3$ hosts channels of both chiralities, partitioning in the injected currents may also occur in the vicinity of $C_S$ and $C_{D}$ and, contrary to Sec.~\ref{sec:s_113}, we should also take into account this contribution. We shall however begin with computing the contribution from the hot spots $C$ and $D$. We proceed as in the previous calculation and use the same conventions. Here, however, the external heat currents are diffusive, while the internal ones are ballistic (see Fig.~\ref{fig:Heat23}). We therefore obtain
\begin{subequations}
\label{eq:23HeatCurrents}
\begin{align}
	&J_{S} = \frac{l_{\rm eq}\kappa T_C^2}{\gamma L_{\rm{Arm}}}, \\
	&J_{u} = \frac{\kappa}{2} T^2_C, \\
	&J_{D1} = \frac{l_{\rm eq}\kappa T_B^2}{ \gamma L_{\rm{Arm}}}, \\
	&J_{G} = \frac{l_{\rm eq}\kappa T_D^2}{ \gamma L_{\rm{Arm}}}, \\
	&J_{d} = \frac{\kappa}{2}T^2_D, \\
	&J_{D2} = \frac{l_{\rm eq}\kappa T_A^2}{\gamma L_{\rm{Arm}}}, \\
	&J_{l} = \frac{\kappa}{2} T^2_A,\\
	&J_{r} = \frac{\kappa}{2} T^2_B,
\end{align}
\end{subequations}
where we have used the same diffusion length $2 l _{\rm eq}/\gamma$ as in the previous section.
Solving again the system of equations \eqref{eq:HeatEqSys1} but now with 
$$P_{A} = 	P_{B} = \frac{e^2}{h}\frac{V_{0}^2}{12}$$ 
and with the heat currents in Eq.~\eqref{eq:23HeatCurrents}, we obtain
\begin{equation}
\label{eq:23TCD}
	k_B^2T^2_C = k_B^2T^2_D = \frac{e^2V_0^2\gamma^2L^2_{\rm{Arm}}}{8\pi^2l_{\rm eq}(l_{\rm eq}+\gamma L_{\rm Arm})} \approx \frac{e^2V_0^2\gamma L_{\rm{Arm}}}{8\pi^2l_{\rm eq}}.
\end{equation}
 Equation (\ref{eq:23TCD}) indicates that the temperature at the noise spots $C$ and $D$ grows with increasing length $L_{\rm Arm}$. The reason for this is that the generated heat from the current injection is due to ballistic charge transport which is length-independent. The generated heat can however only leave the QPC region by diffusion. The larger length segments $L_{\rm Arm}$, the less heat leaves the system. Hence, the steady state results from constant heating that slowly diffuses away. In practice, if $L_{\rm Arm}$ is made larger, effects of heat leakage to the bulk will become important at some point, so that the temperature will not increase without bound.

Combining Eqs.~\eqref{eq:23TCD}, \eqref{eq:noiseD1} and~\eqref{eq:FanofactorQPC}, we find
\begin{equation}
	F_{\rm{QPC}} = \frac{k_B(T_C+T_D)}{eV_0} \simeq 0.23 \sqrt{\frac{\gamma L_{\rm{Arm}}}{l_{\rm eq}}}.
\end{equation}
What remains to compute is the contribution $F_{\rm Contact}$ from the noise spots $E$ and $F$. The microscopic structure of $\nu_{\rm B}=2/3$ is $\nu_+ = 1$, $\nu_- = 1/3$, $n_u = 1$, and $n_d=1$. To obtain the temperature profiles of the downstream and upstream modes, we solve Eq.~\eqref{eq:TemperatureEquation}, with boundary conditions $T_+(0)=0$ and $T_-(L_{\rm{Arm}}) = T_C$. Again, no voltage drop occurs along the outer edge segments $C_S - C$ and $C_G - D$, and we find the diffusive temperature profiles 
\begin{subequations}
	\begin{align}
		& k_B^2T^2_+(x) \simeq \frac{e^2V_0^2 x \gamma }{8 \pi^2 l_{\rm eq}}, \\
		& k_B^2T^2_-(x) \simeq \frac{e^2V_0^2 (x \gamma+ 2 l_{\rm eq}) }{8 \pi^2 l_{\rm eq}}, 
	\end{align}
\end{subequations}
where we used $l_{\rm eq}\ll L_{\rm Arm}$. We note that the $L_{\rm Arm}$ dependence drops out for these profiles since the 
large factor $\sim L_{\rm Arm}^{1/2}$ originating from the temperature at the noise spot (see Eq.~\eqref{eq:23TCD}) is compensated by a small factor $\sim 1/\sqrt{L_{\rm Arm}}$ describing the reduction of the temperature towards the external contact. 
The corresponding contribution to the noise should therefore stay constant in the limit $l_{\rm eq}\ll L_{\rm Arm}$. Indeed, plugging these profiles into Eq.~\eqref{eq:NoiseExp2}, we obtain the asymptotically constant noise
\begin{align}
\label{eq:hgamma}
	&\overline{(\delta I_S)^2} = \overline{(\delta I_G)^2} \simeq \frac{e^3V_0}{h} \frac{\sqrt{\gamma}\left( \sqrt{\pi}+2e^{4/\gamma}\Gamma(\frac{3}{2},\frac{4}{\gamma}) \right) }{36\pi} \notag \\
	&  \equiv \frac{e^3 V_0}{h} h(\gamma).
\end{align}
Inserting this result into Eq.~\eqref{eq:FanofactorQPC}, we get
\begin{equation}
	F_{\rm Contact} = 3 h(\gamma),
\end{equation}
which yields the final expression for the total Fano factor
\begin{equation}
	F_{2/3,1/3} \simeq 0.23 \sqrt{\frac{\gamma L_{\rm{Arm}}}{l_{\rm eq}}}+3h(\gamma).
	\label{eq:F2313}
\end{equation}
This result is in agreement with qualitative discussion in Sec.~\ref{sec:QuantNoise} (see Eq.~\eqref{eq:diffusivenoise1}). 
The Fano factor in this case increases as $(L_{\rm{Arm}} /  l_{\rm eq})^{1/2}$ 
 because the only way the heat generated at the hot spots can exit the QPC region is by diffusion. This leads to a temperature at the noise spots that rises with increasing $L_{\rm Arm}$. Since the noise measured in the drains is proportional to the temperature at the noise spot, the asymptotics of the Fano factor follows. 
 
 As has already been discussed above, for very large $L_{\rm{Arm}}$, the increase of the noise will be limited by the leakage of heat to the bulk. Specifically, if $L_{\rm{Arm}}$ becomes larger than the leakage length, it will be replaced by the leakage length in the first (dominant) term of Eq.~\eqref{eq:F2313}. The contact contribution (second term of Eq.~\eqref{eq:F2313}) will drop out in this situation. 

%

\end{document}